\pdfoutput=1
\newcommand*{\ATLASLATEXPATH}{latex/}
\documentclass[cernpreprint, texlive=2016, UKenglish, LANGEDIT=false]{\ATLASLATEXPATH atlasdoc}
\usepackage{\ATLASLATEXPATH atlaspackage}
\usepackage{\ATLASLATEXPATH atlasbiblatex}
\usepackage{\ATLASLATEXPATH atlasphysics}
\usepackage{layout}
\graphicspath{{logos/}{figures/}}
\usepackage{ttbar_bjet_calibrations_2015-defs}

 \AtlasTitle{Measurements of $b$-jet tagging efficiency with the ATLAS detector using \ttbar\ events at $\sqrt{s}=13$~TeV}
 \author{The ATLAS Collaboration}
 \AtlasRefCode{PERF-2016-05}
 \AtlasJournal{JHEP}
 \AtlasAbstract{%
The efficiency to identify jets containing $b$-hadrons ($b$-jets) is measured using a high purity sample of dileptonic top quark--antiquark pairs ($t\bar{t}$) selected from the
36.1~fb$^{-1}$ of data collected by the ATLAS detector in 2015 and 2016 from proton--proton collisions produced by the Large Hadron Collider at a centre-of-mass energy
$\sqrt{s}=13$~TeV.  Two methods are used to extract the
efficiency from $t\bar{t}$ events, a combinatorial likelihood
approach and a tag-and-probe method.
A boosted decision tree, not using $b$-tagging information, is used to select events in which
two $b$-jets are present, which reduces the dominant
uncertainty in the modelling of the flavour of the jets. 
The efficiency is extracted for
jets in a transverse momentum range from 20 to 300~GeV, with
data-to-simulation scale factors calculated by comparing the
efficiency measured using collision data to that predicted by the
simulation. The two methods give compatible results, and achieve a similar level of precision,
measuring data-to-simulation scale factors close to unity with
uncertainties ranging from
2\% to 12\% depending on the jet transverse momentum.
}

\PreprintIdNumber{CERN-EP-2018-047}
\AtlasJournalRef{JHEP 08 (2018) 89}
\AtlasDOI{10.1007/JHEP08(2018)089}

\hypersetup{pdftitle={ATLAS document},pdfauthor={The ATLAS Collaboration}}

\begin{document}

\LEcontact{Pippa Wells pippa.wells@cern.ch}

\maketitle

\tableofcontents
\clearpage

\section{Introduction}
\label{sec:intro}

The identification of jets containing $b$-hadrons, referred to as \textit{$b$-jets}, is vital for a large part of the physics programme of the ATLAS experiment~\cite{Aad:2008zzm} at the CERN Large Hadron Collider (LHC), including Standard Model (SM) precision measurements, studies of the Higgs boson's properties and searches for new physics beyond the SM. The algorithms used to identify \bjets\ are referred to as \textit{$b$-tagging} algorithms.
 
This paper describes a measurement of the $b$-jet tagging efficiency in proton--proton collision data recorded at \com{13}\ during Run~2 of the LHC.   A very pure \ttbar\ sample is selected, as these events have a high $b$-jets purity by virtue of the $t \rightarrow Wb$ branching fraction being close to 100\%~\cite{Patrignani:2016xqp}. The number of additional non-$b$-jets in the sample is greatly reduced by requiring that both $W$ bosons decay leptonically.
Two methods are used to measure the \beff: a new method which uses a tag-and-probe approach, referred to as the \textit{Tag-and-Probe} method (T\&P); and a combinatorial likelihood approach, referred to as the \textit{Likelihood} method (LH), which is based upon a method used during Run~1 (\com{7}\ and \com{8}) of the LHC~\cite{PERF-2012-04}. Having two methods enables reciprocal cross-checks to be made between them.

The $b$-jet tagging efficiency, $\varepsilon_{b}$, is measured for jets in the pseudorapidity\footnote{ATLAS uses a right-handed coordinate system with its origin at the nominal interaction point (IP) in the centre of the detector and the $z$-axis along the beam pipe. The $x$-axis points from the IP to the centre of the LHC ring, and the $y$-axis points upward. Cylindrical coordinates $(r,\phi)$ are used in the transverse plane, $\phi$ being the azimuthal angle around the $z$-axis. The pseudorapidity is defined in terms of the polar angle $\theta$ as $\eta=-\ln\tan(\theta/2)$. The angular distance $\Delta R$ is measured in $\eta$--$\phi$ phase space and is defined as $\sqrt{(\Delta\eta)^{2} + (\Delta\phi)^{2}}$, where $\Delta\eta$ and $\Delta\phi$ are the differences between the $\phi$ and $\eta$ of the two objects respectively.} range $|\eta|<$~2.5 and with transverse momentum \pt\ $>$ 20~GeV for several \textit{operating points} (OP). Operating points are defined by sets of selection criteria imposed upon the output of the $b$-tagging algorithm designed to provide a certain \beff. Four operating points are defined, corresponding to 60\%, 70\%, 77\% and 85\% $b$-jet tagging efficiencies in simulated \ttbar\ events. Two sets of four operating points are implemented to provide a \textit{single-cut} or a \textit{flat-efficiency} operating point. The \textit{single-cut} operating point provides the stated \beff\ when averaged over the transverse momentum distribution of \bjets\ in \ttbar\ events, but the efficiency varies with jet~\pt. On the other hand, the \textit{flat-efficiency} operating point has a varying cut value, ensuring a constant \beff\ as a function of the jet \pt. Results are also presented in the form of data-to-simulation efficiency scale factors, defined as ${\varepsilon_{b}^\mathrm{data}}/{\varepsilon_{b}^\mathrm{sim}}$,
where $\varepsilon_{b}^\mathrm{data}$ is the efficiency measured in data, while ${\varepsilon_{b}^\mathrm{sim}}$ represents the efficiency predicted by simulation using Monte Carlo (MC) generator-level information. In physics measurements, these scale factors can be applied jet by jet to correct the rate of events after applying a $b$-tagging requirement.
The scale factors are measured for all operating points; however, this paper presents only results from a number of selected working points as examples.
 Separate measurements have also been made for the tagging efficiencies of jets containing $c$-hadrons, referred to as \textit{$c$-jets}, and for jets containing neither a $b$-hadron nor a $c$-hadron, referred to as \textit{light-flavour} jets, and are presented in Ref.~\cite{PERF-2012-04}. 

The paper is organised as follows. In Section~\ref{sec:detector}, the ATLAS detector and physics object reconstruction are described. Section~\ref{sec:algs} contains a description of the ATLAS $b$-tagging algorithms. In Section~\ref{sec:samples}, the data and simulated samples used in the $b$-jet tagging efficiency measurements are presented. Section~\ref{sec:CommonSelec} summarises the event selection criteria applied for both calibration methods, while in Section~\ref{sec:methods} the T\&P and LH methods are presented in detail. In Section~\ref{sec:systematics}, the systematic uncertainties for each method are outlined, and results are presented in Section~\ref{sec:results}.  Finally, conclusions are given in Section~\ref{sec:conclusion}.


\section{The ATLAS detector and object reconstruction}
\label{sec:detector}


The ATLAS detector~\cite{Aad:2008zzm} at the LHC covers nearly the entire solid angle around the collision point. The detector comprises an inner tracking detector surrounded by a superconducting solenoid producing a 2~T axial magnetic field, a system of calorimeters, and a muon spectrometer (MS) incorporating three large toroid magnet assemblies.  The inner detector (ID) consists of four layers of silicon pixel sensors and four layers of silicon microstrip sensors, 
providing precision tracking in the pseudorapidity range $|\eta| < 2.5$. The innermost pixel layer, referred to as the insertable B-layer (IBL)~\cite{Capeans:1291633,Abbott:2018ikt}, was installed between Run~1 and Run~2 of the LHC.
The IBL provides a hit measurement at an average radius of 33.3~mm, significantly closer to the interaction point than the closest pixel layer in Run~1 (radius of 50.5 mm).  The additional pixel layer has a significant impact on the 
performance of both the tracking and vertexing algorithms, resulting in improved $b$-tagging performance. 
  A straw-tube transition radiation tracker complements the measurements in the silicon layers by providing additional tracking and electron identification information for $|\eta| < 2.0$.

High-granularity electromagnetic (EM) and hadronic sampling calorimeters cover the region $|\eta| < 4.9$. All electromagnetic calorimeters, as well as the endcap and forward hadronic calorimeters, use liquid argon as the active medium and lead, copper, or tungsten absorber. The central hadronic calorimeter uses scintillator tiles as the active medium and steel absorber. The muon spectrometer measures the deflection of muons with $|\eta| < 2.7$ using multiple layers of high-precision tracking chambers located in a toroidal field of approximately 0.5~T or 1~T in the central and endcap regions of ATLAS, respectively. 

The ATLAS detector incorporates a two-level trigger system, with the first level implemented in custom hardware and the second level implemented in software. This trigger system reduces the output from the detector electronics to about 1~kHz for offline storage.

Vertices are reconstructed using tracks measured  by the inner detector~\cite{Aaboud:2016rmg}.  Events are required to have at least one reconstructed vertex, with two or more associated
tracks which have $p_{\text{T}}>$~400~MeV. The primary vertex
is chosen as the vertex candidate with the largest
sum of the squared transverse momenta of associated tracks.

Electrons are reconstructed from energy deposits (clusters) in the electromagnetic calorimeter matched to tracks reconstructed in the ID~\cite{PERF-2013-03,ATLAS-CONF-2016-024}. Additionally, candidate clusters in the calorimeter barrel/endcap transition region, defined by $1.37 < |\eta_{\textrm{cluster}}| < 1.52$, as well as those of poor quality, are excluded.
Muons are reconstructed from track segments in the MS that are matched to tracks in the ID~\cite{Aad:2014rra,PERF-2015-10}. Combined muon tracks are then re-fit using information from both the ID and MS systems.
 The lepton tracks must be consistent with coming from the primary vertex of the event: the longitudinal impact parameter $z_{0}$ must satisfy $|z_{0} \sin{\theta}| < 0.5$~mm, while the transverse impact parameter significance, $|d_{0}|/\sigma (d_{0})$ must be less than 5 for electrons or less than 3 for muons. To reduce the contribution from hadronic decays (non-prompt leptons), photon conversions and hadrons misidentified as leptons, both the electrons and muons must also satisfy isolation and identification criteria.  The loose, medium and tight working points of the isolation and identification algorithms are defined in Ref.~\cite{ATLAS-CONF-2016-024} for electrons, and in Ref.~\cite{PERF-2015-10} for muons. Two types of leptons are defined for the analyses presented in this paper. First, \textit{signal} lepton candidates are required to have \pT\ > 27~GeV and $|\eta| < 2.5$, as well as to satisfy tight track- and calorimeter-based isolation criteria.~\textit{Signal} electrons (muons) are required to pass the medium electron (muon) identification criteria. Second, \textit{loose} leptons are required to have \pT~>~7~GeV and $|\eta| < 2.5$, as well as to satisfy loose identification and loose track-only isolation criteria.

Jets are reconstructed from three-dimensional topological energy clusters~\cite{PERF-2014-07} in the calorimeter using the \antikt\ algorithm~\cite{Cacciari:2008gp} with a radius parameter of $R = 0.4$. These jets are referred to as \textit{calorimeter-jets}. The clusters are calibrated to the electromagnetic energy response scale prior to jet reconstruction. The reconstructed jets are then calibrated to the jet energy scale (JES), corresponding to the particle scale\footnote{The particle scale is defined as consisting of stable particles emerging from the p-p collision before interaction with the ATLAS detector.}, using corrections derived from simulation and \textit{in situ} corrections based on 13~TeV data~\cite{Aaboud:2017jcu}. Jets are required to have calibrated \pT\ > 20~GeV and to be within the acceptance of the inner detector, $|\eta| < 2.5$.  Jet cleaning criteria are applied to identify jets arising from non-collision sources or noise in the calorimeter~\cite{ATLAS-CONF-2012-020,ATLAS-CONF-2015-029}.  Any event containing such a jet is removed. In order to reduce the contamination from jets arising from additional $pp$ collisions in the same or nearby bunch crossings, called pile-up, a requirement on the Jet Vertex Tagger (JVT)~\cite{PERF-2014-03} output is made.  The JVT algorithm combines tracking information into a multivariate algorithm to reject jets which do not originate from the primary vertex, and is applied to jets with \pT\ < 60~GeV and $|\eta| < 2.4$.  Jets with  \pT\ > 60~GeV are assumed to have originated from the primary vertex.

Jets are also reconstructed from inner-detector tracks using the \antikt\ algorithm with a radius parameter of $R = 0.2$.  These jets are referred to as \textit{track-jets}. The tracks used in jet clustering are required to have~\mbox{\pT\ > 0.5~GeV} and to be matched to the primary vertex using impact parameter requirements on the tracks.
Only track-jets with at least two tracks and with \pT\ > 10~GeV and $|\eta| < 2.5$ are considered for the purposes of the $b$-jet tagging efficiency measurement. The results presented in this paper correspond to the jets reconstructed from the topological energy clusters in the calorimeter, which are referred to as \textit{jets} throughout.  Equivalent $b$-jet efficiency measurements are also performed for track-jets and the results made available to ATLAS analyses using those jets.

In order to avoid counting a single detector response as originating from two different objects, an overlap removal procedure is applied to the jet candidates and leptons passing the loose quality requirement. To prevent double-counting of electron energy deposits reconstructed as jets, the closest jet lying $\Delta R < 0.2$ from a selected electron is removed.
Electron candidates that lie  $\Delta R < 0.4$ from a jet surviving the selection are discarded to reduce the background from electrons that originate from heavy-flavour decays.
 Furthermore, to reduce the background from muons that originate from the decays of hadrons containing a heavy quark inside selected jets, muon candidates are removed if they are separated from the nearest selected jet by $\Delta R < 0.4$. However, if this jet has fewer than three associated tracks, the muon is kept and the jet is removed as it is likely that the energy is deposited in the calorimeter by the muon.

The missing transverse momentum (with magnitude \met) is defined as the negative vector sum of the transverse momenta of all selected and calibrated physics objects in the event, with an extra ``soft'' term added to account for low-momentum contributions from particles in the event that are not associated with any of the selected objects. This term is calculated using inner-detector tracks matched to the primary vertex to reduce the pile-up contamination~\cite{Aaboud:2018tkc}.

\section{Definition of $b$-tagging algorithms} \label{sec:algs}

A new multivariate $b$-tagging algorithm, referred to as MV2c10, was developed for Run~2, and utilises a boosted decision tree (BDT). The algorithm is similar to the multivariate algorithms developed during Run~1~\cite{PERF-2012-04}, but with a dedicated optimisation carried out for Run~2 to exploit the installation of the IBL and improved tracking software~\cite{ATL-PHYS-PUB-2015-022, ATL-PHYS-PUB-2016-012}. The algorithms which provide the input variables for MV2c10 all exploit the relatively long $b$-hadron lifetime: a likelihood-based combination of the transverse and longitudinal impact parameter significances; the presence of a secondary vertex and related properties; and the reconstruction of the $b$-hadron decay chain using a Kalman filter to search for a common direction connecting the primary vertex to both the bottom and the tertiary charm decay vertices. Additionally, the jet \pt\ and jet $\eta$ are included as BDT training variables to take advantage of correlations with other variables. In order to avoid any difference between the kinematic spectra of \bjets\ and background jets being used as a discriminating variable, the $b$-jet \pt\ and $\eta$ spectra are reweighted to match the combined $c$-jet and light-flavour jet spectrum.  The BDT was trained on a subset of events from a simulated \ttbar\ sample, produced with \textsc{Powheg}~\cite{Nason:2004rx,Frixione:2007vw,Alioli:2010xd,Powhegttbar} interfaced with \textsc{Pythia6} for the parton shower, hadronisation, and the underlying event~\cite{Sjostrand:2006za} and using the CT10~\cite{Lai:2010vv} parton distribution function set, as described in more detail in Section~\ref{sec:samples}.  The BDT training is performed by assigning \bjets\ as signal, and \cjets\ and light-flavour jets as background.  In order to enhance the \cjet\ rejection, the \cjet\ fraction in the training is set to 7\%, and the light-flavour jet background is set to 93\%, as described in Ref.~\cite{ATL-PHYS-PUB-2016-012}. 

\begin{figure}[htb!]
\centering
\subfloat[][]{
\includegraphics[width=0.50\textwidth]{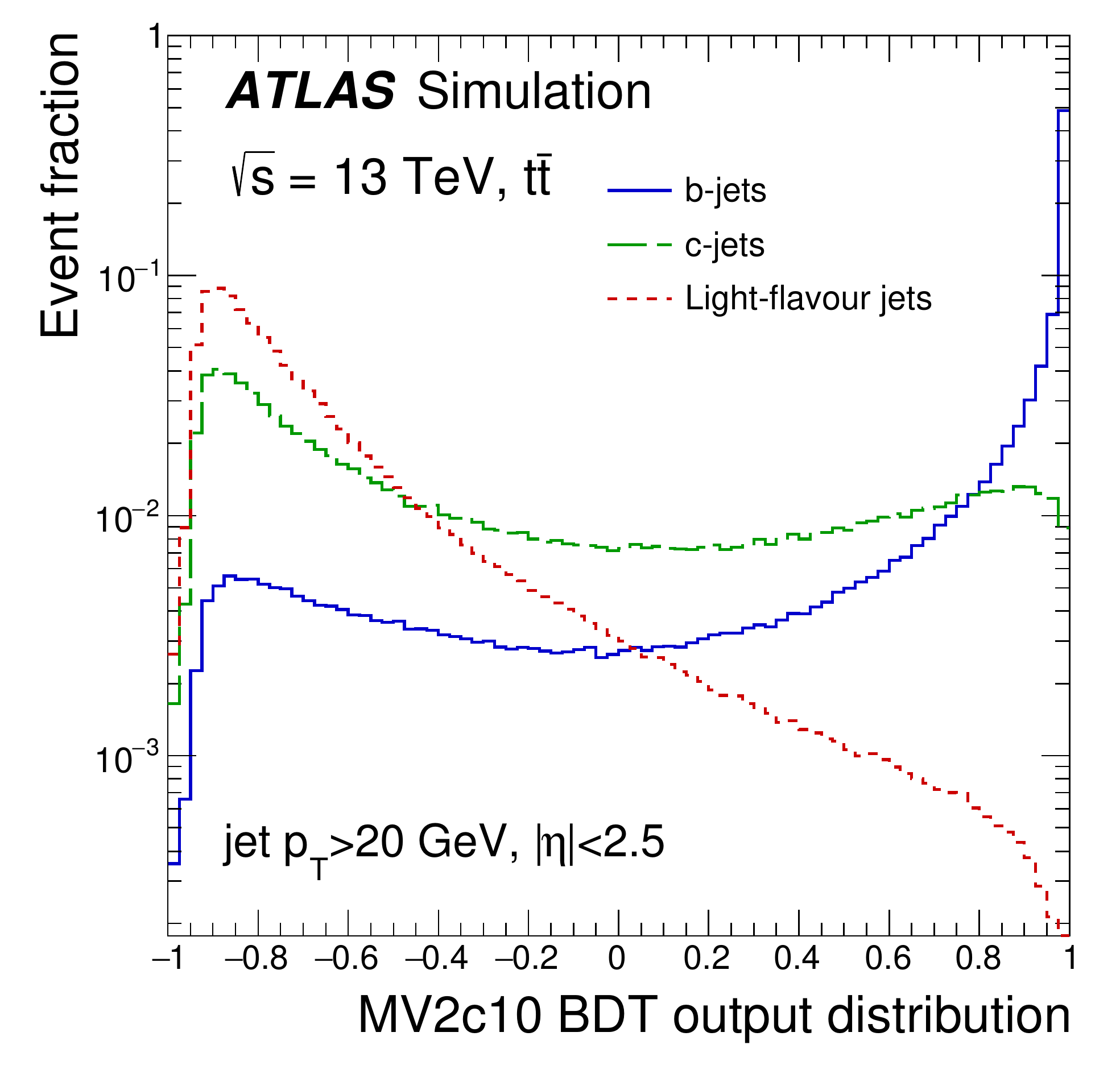}
\label{fig:mv2output}
}
\subfloat[][]{
\includegraphics[width=0.50\textwidth]{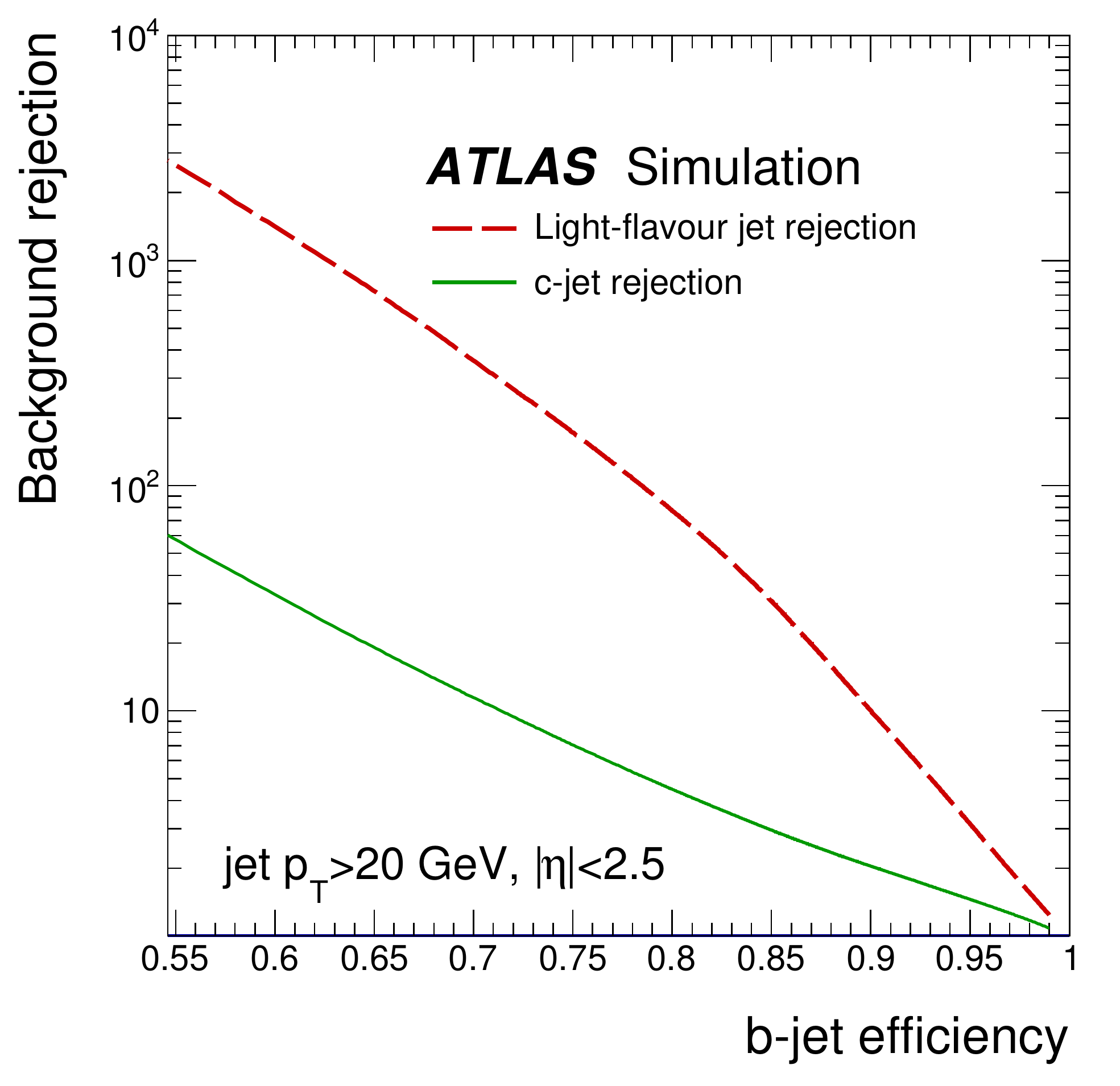}
\label{fig:theCompmv1c}
}
\caption{(a) The MV2c10 output for $b$-jets (solid line), $c$-jets (dashed line) and light-flavour jets (dotted line) in simulated $t\bar{t}$ events. (b) The light-flavour jet (dashed line) and $c$-jet rejection factors (solid line) as a function of the $b$-jet tagging efficiency of the MV2c10 $b$-tagging algorithm. The performance was evaluated on $t\bar{t}$ events simulated using \textsc{Powheg} interfaced to \textsc{Pythia6}.}
\label{fig:mv2}
\end{figure}

\begin{figure}[htb!]
\centering
\subfloat[][]{
\includegraphics[width=0.50\textwidth]{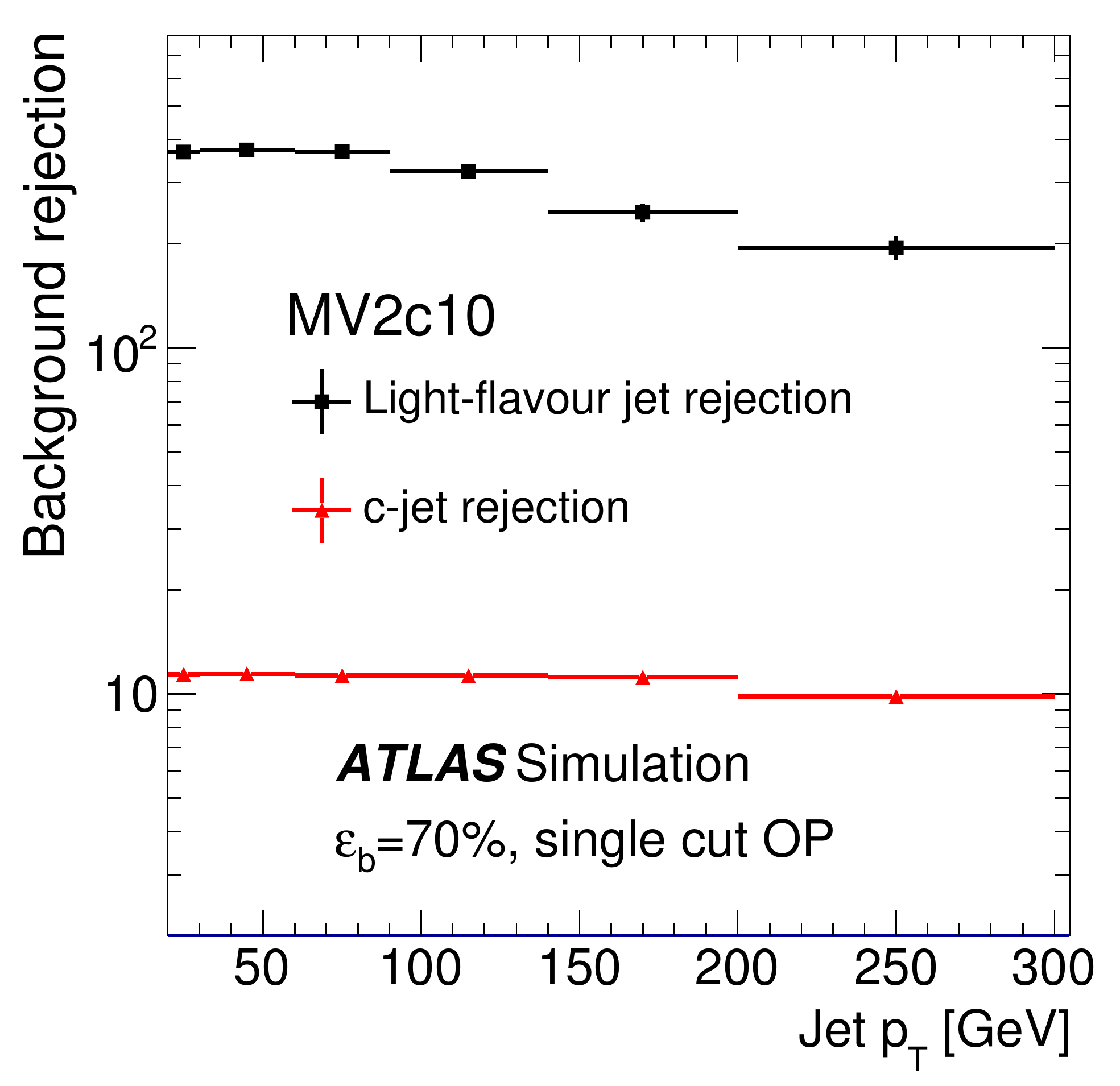}
\label{fig:rejectionmv2c10singCut}
}
\subfloat[][]{
\includegraphics[width=0.50\textwidth]{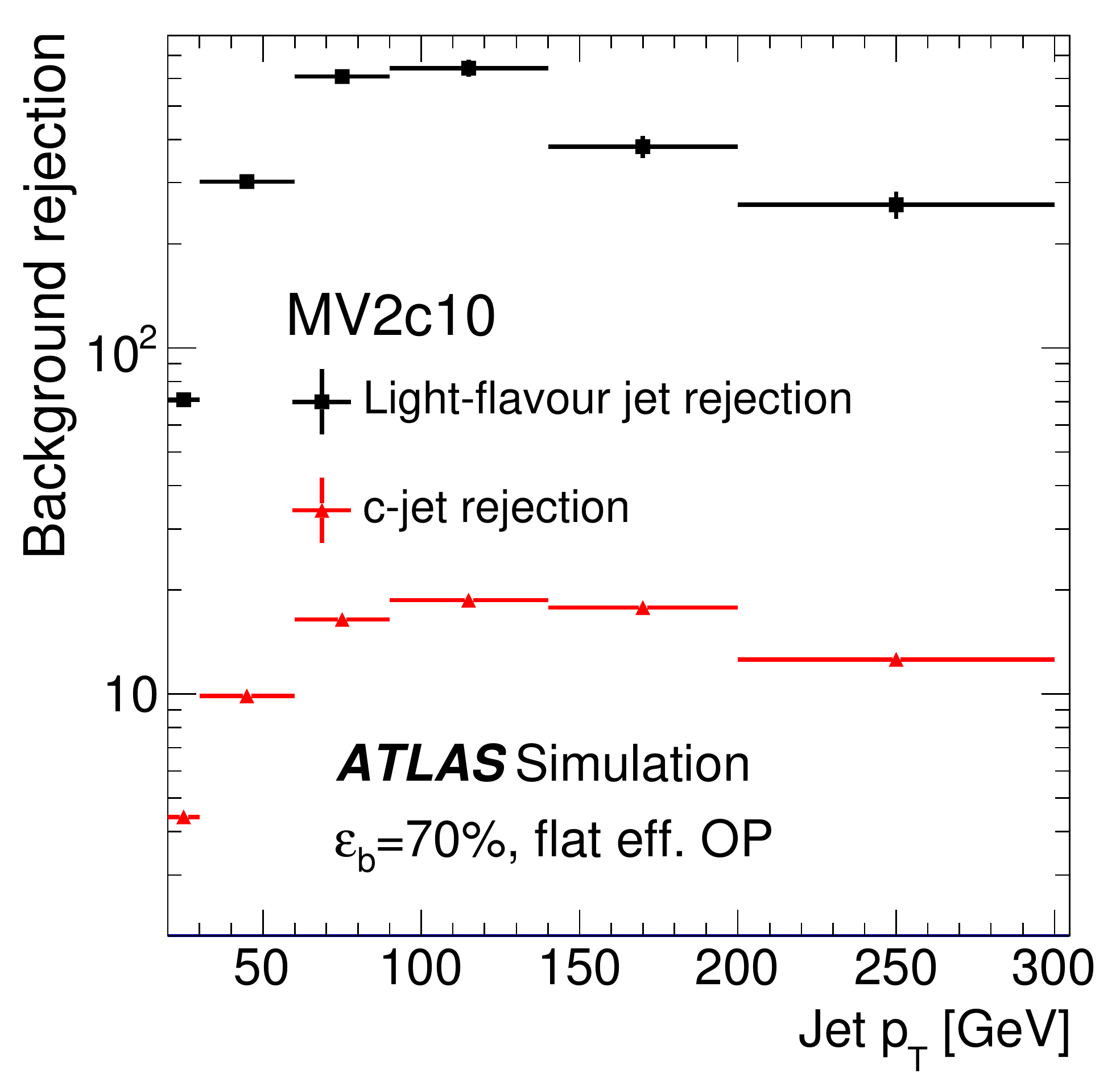}
\label{fig:rejectionmv2c10flat}
}
\caption{The light-flavour jet (squares) and $c$-jet rejection factors (triangles) at a $b$-tagging efficiency of 70\% corresponding to (a) the single-cut OP and (b) the flat-efficiency OP as a function of the jet \pt\ for the MV2c10 $b$-tagging algorithms in $t\bar{t}$ events simulated using \textsc{Powheg} interfaced to \textsc{Pythia6}.}
\label{fig:rejectionmv2c10}
\end{figure}

The MV2c10 output for $b$-jets, $c$-jets and light-flavour jets in a \ttbar\ sample, which is statistically independent from the training sample, is presented in Figure~\ref{fig:mv2output}.  
The rejection rates for light-flavour jets and $c$-jets are defined as the inverse of the efficiency for tagging a light-flavour jet or a $c$-jet as a \bjet, respectively. Figure~\ref{fig:theCompmv1c} shows the corresponding light-flavour jet and $c$-jet rejection factors as a function of the $b$-jet tagging efficiency. The rejection rates for both the light-flavour jets and $c$-jets as a function of jet \pt\ are given in Figure~\ref{fig:rejectionmv2c10singCut} for the single-cut OP and Figure~\ref{fig:rejectionmv2c10flat} for the flat-efficiency OP, both for a 70\% $b$-jet tagging efficiency.

\section{Dataset and simulated event samples}
\label{sec:samples}


The data used in these measurements were collected by the ATLAS
detector from proton--proton collisions in 2015 and 2016 at a centre-of-mass energy $\sqrt{s}=13$~TeV with 25~ns proton bunch spacing. The data correspond to a
total integrated luminosity of $36.1 \pm 1.2$~fb$^{-1}$ after offline data
quality selection, measured following Ref.~\cite{Aaboud:2016hhf}.

The dominant \ttbar\ process was modelled using the
matrix-element generator \textsc{Powheg-Box
  v2}~\cite{Nason:2004rx,Frixione:2007vw,Alioli:2010xd,Powhegttbar}, which provides
next-to-leading-order (NLO) accuracy in QCD. It used the CT10 parton distribution function (PDF)
set~\cite{Lai:2010vv} and it was interfaced to \textsc{Pythia 6.428}~\cite{Sjostrand:2006za} with the Perugia 2012 set~\cite{Skands:2010ak} of tuned parameters (tune) for the modelling of the parton shower, fragmentation and the underlying event. The $h_\mathrm{damp}$ parameter,
which controls the \pt\ of the first additional emission beyond the
Born configuration, was set to $m_{t} = 172.5$~GeV, a setting that was
found to improve the description of the \pt\ of the \ttbar\ system when compared to data~\cite{ATL-PHYS-PUB-2016-004}.

The dominant non-\ttbar\ process is the associated production of a single top quark and a $W$ boson ($Wt$ process), which also contains a large fraction of $b$-jets.
 Other processes such as \zjets\ and diboson production
constitute only a small fraction of the sample. Table~\ref{tab:MC}
summarises the nominal MC generators used to simulate physics processes, along with the alternative samples used to
estimate the systematic uncertainties related to the choice of MC generator and associated parameter settings.  The contamination from $t\bar{t}$ produced in association
with either a vector boson or a SM Higgs boson, and the gluon--gluon fusion or vector-boson fusion production of a Higgs boson,
which
subsequently decays into a pair of $W$ bosons, were found to be negligible. Therefore, these processes are not
considered further.  

Events in which one of the two selected lepton candidates is not a real prompt lepton (e.g.~one coming from a $b/c$~hadron decay, a photon conversion, or a hadron misidentified as a lepton) are referred to as \text{misidentified} lepton events, and are estimated from data by changing the selection criterion from opposite-charge to same-charge leptons.  In order to avoid double counting the contribution of misidentified leptons which is already reproduced by Monte Carlo, and to partially take into account a possible difference between opposite-charge and same-charge lepton events,
the contribution from simulated same-charge events is subtracted from data for the estimate of the misidentified lepton background.

\begin{table}[htb]
\renewcommand{\arraystretch}{1.1} 
\centering
\begin{footnotesize}
\centering
\begin{tabular}{>{\centering\arraybackslash}p{2.20cm}|>{\centering\arraybackslash}p{3.00cm}|>{\centering\arraybackslash}p{3.00cm}|>{\centering\arraybackslash}p{3.05cm}|>{\centering\arraybackslash}p{2.0cm}}
 \hline 
 \hline 
 Process & Generator & PDF Set, Tune & {Hadronisation/ Fragmentation}
  & Order in pQCD  of Inclusive $\sigma$
 \\ \hline  \hline 
$t\bar{t}$ & \textsc{Powheg-Box v2} & \textsc{CT10, Perugia2012} & \textsc{ Pythia-6.428} & NNLO+NNLL~\cite{Czakon:2011xx} \\ 
Single top ($Wt$) &\textsc{Powheg-Box v1}& \textsc{CT10, Perugia2012} & \textsc{ Pythia-6.428} & NNLO~\cite{Kidonakis:2010ux} \\
  $Z/ \gamma^{\ast}$ + jets &  \textsc{MG5\_aMC@NLO 2.2.2~\cite{Alwall:2014hca,Lonnblad:2001iq}} &\textsc{NNPDF23LO~\cite{NNPDF}, A14~\cite{ATL-PHYS-PUB-2014-021}} & \textsc{Pythia-8.186}~\cite{Sjostrand:2006za,Pythia8} & NNLO~\cite{Catani:2009sm}\\
 Diboson &\textsc{Sherpa 2.1.1}~\cite{sherpa} & \textsc{CT10 }  & \textsc{Sherpa 2.1.1} & NLO  \\
 \hline
\hline
\multicolumn{5}{c}{Alternative Generators}\\
\hline
\hline
$t\bar{t}$ & \textsc{Powheg-Box v2} & \textsc{CT10, Perugia2012 radLo~\cite{ATL-PHYS-PUB-2015-002}} & \textsc{ Pythia-6.428} & 
NNLO+NNLL \\ 
$t\bar{t}$ & \textsc{Powheg-Box v2} & \textsc{CT10, Perugia2012 radHi~\cite{ATL-PHYS-PUB-2015-002}} & \textsc{ Pythia-6.428} & 
NNLO+NNLL \\ 
$t\bar{t}$ & \textsc{Powheg-Box v2} & \textsc{CT10, UE-EE-5} & \textsc{Herwig++ 2.7.1}~\cite{herwigpp} & 
NNLO+NNLL \\ 
$t\bar{t}$ & \textsc{MG5\_aMC@NLO 2.2.2} & \textsc{CT10, UE-EE-5~\cite{ATL-PHYS-PUB-2014-021}} & \textsc{ Herwig++ 2.7.1} & 
NNLO+NNLL \\ 
Single top &\textsc{Powheg-Box v1}& \textsc{CT10, Perugia2012 radLo} & \textsc{ Pythia-6.428} & 
NNLO \\
Single top &\textsc{Powheg-Box v1}& \textsc{CT10, Perugia2012 radHi} & \textsc{ Pythia-6.428} & 
NNLO \\
Single top &\textsc{Powheg-Box v1}& \textsc{CT10, UE-EE-5} & \textsc{Herwig++ 2.7.1} & 
NNLO \\
Single top & \textsc{MG5\_aMC@NLO 2.2.2} & \textsc{CT10, UE-EE-5~\cite{ATL-PHYS-PUB-2014-021}} & \textsc{ Herwig++ 2.7.1} & 
NNLO+NNLL \\
   Single top &\textsc{Powheg-Box v1}& \textsc{CT10, Perugia2012} & \textsc{ Pythia-6.428} & NNLO \\
 $Z/ \gamma^{\ast}$ + jets & \textsc{Powheg-Box v2}  &\textsc{CT10, AZNLO~\cite{Aad:2014xaa}} & \textsc{ Pythia-8.186} & NNLO \\
\hline
\hline

\end{tabular}
\end{footnotesize}
\caption{
A summary of Monte Carlo generators used to simulate various physics processes, together with their basic parameter settings and corresponding cross-section order in pQCD at \com{13}.  
  Whenever  \textsc{ Pythia} or \textsc{Herwig++}
is used for parton shower simulation, the parton shower PDFs are taken
from \textsc{CTEQ6L1}.}
\label{tab:MC}
\end{table}

The \textsc{EvtGen}~\cite{Lange:2001uf} package was used with all the hadronisation/fragmentation generators, except for \textsc{Sherpa}, to model the decays of $b$- and $c$-hadrons. All simulated samples were processed through
the ATLAS detector simulation~\cite{SOFT-2010-01} based on
GEANT4~\cite{Agostinelli:2002hh}.
Additional simulated $pp$ collisions
generated with \textsc{Pythia8}~\cite{Pythia8} were overlaid on all simulated samples to model the expected number of additional pile-up interactions in each event.
Simulated events are corrected so that the lepton and jet identification
efficiencies, energy scales and energy resolutions match those
determined from data control samples at $\sqrt{s}$~=~13~TeV.

In the simulated samples, 
a reconstructed jet is labelled as a \bjet\ if, within $\Delta R = 0.3$, there is a matching weakly decaying \bhadron\ with \pT\ > 5~GeV.
The flavour labelling is exclusive, with
the hadron matched to the closest jet in the $\Delta R$ phase space.  If no \bhadron\
is found, but a \chadron\ is matched to the jet, then it is labelled as
a \cjet. If there is no $b$- or \chadron, but a $\tau$-lepton is matched to the jet, then it is labelled as a $\tau$-jet, otherwise it is labelled as a light-flavour jet.


\section{Event selection}
\label{sec:CommonSelec}
Events were recorded using triggers requiring at least one lepton, with lepton isolation requirements and \pt\ thresholds that vary depending on the
data-taking conditions. In 2015 this threshold was 20~GeV for muons
and 24~GeV for electrons, while in 2016 it was raised to 26~GeV for
both the electrons and muons.
These triggers
are combined with higher-threshold triggers, of 50~GeV for muons and
60, 120 and 140~GeV for electrons, without isolation requirements, to improve the
trigger efficiency for leptons with high transverse momentum. 

\begin{figure}
\begin{center}
\subfloat[][]{
\includegraphics[width=0.50\textwidth]{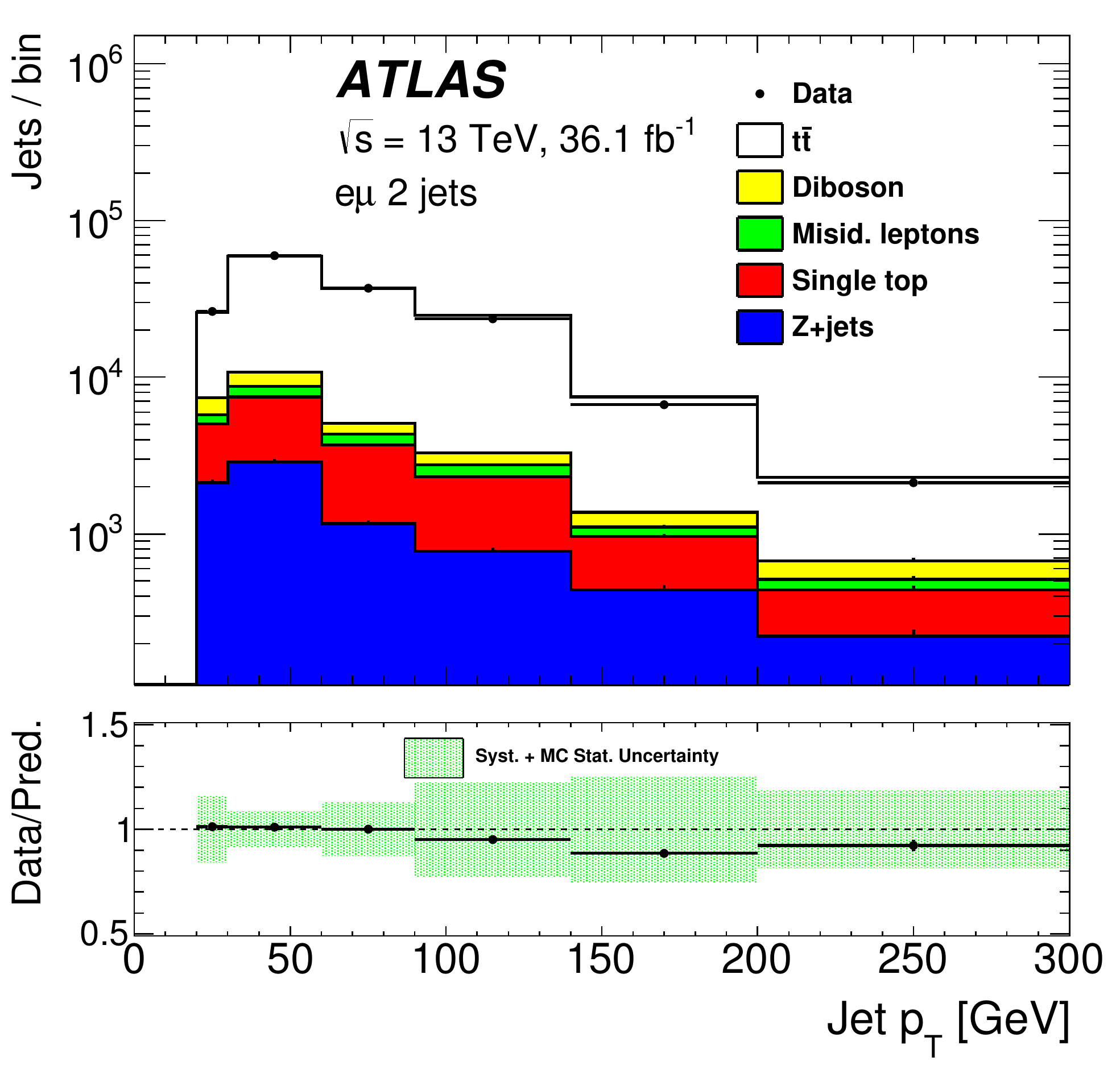}\label{fig:PDFmodellingNoBDT_a}}
\subfloat[][]{
\includegraphics[width=0.50\textwidth]{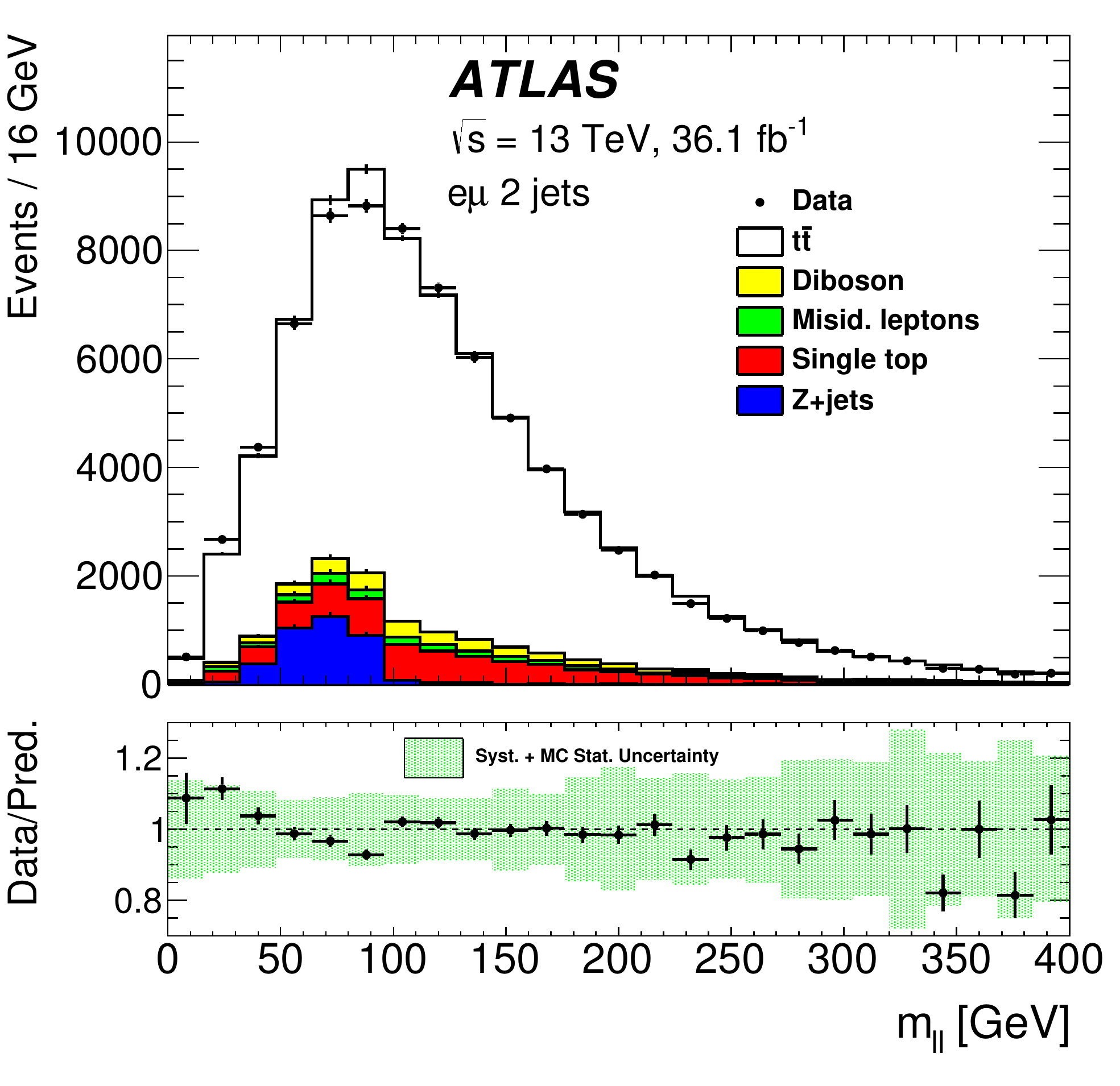}\label{fig:PDFmodellingNoBDT_b}}
\caption{The (a) jet \pt\ distribution and (b) $m_{\ell\ell}$ distribution
  for the data (points) and simulated samples (stacked
  histograms) for the $e\mu+$ 2-jet selection in the LH method. The simulated samples are
  normalised to agree with a fit to the data as described in
  Section~\ref{subsec:PDFmethod}. The bottom panels show the ratio of
  the data to the simulated samples, with the dotted uncertainty
  bands corresponding to the total MC statistical uncertainty and systematic uncertainty.}
  \label{fig:PDFmodellingNoBDT}
\end{center}
\end{figure}

Table~\ref{tab:Selections} summarises the event selection criteria specific to
the Likelihood and T\&P methods. The events are selected by requiring two oppositely charged signal
leptons ($e\mu$, $ee$ or $\mu\mu$) and two or three jets. One of the leptons must also be matched using a $\Delta R$ requirement to one of the objects that triggered the event. A veto is applied to events
which contain one or more additional loose leptons. 
The T\&P method uses only the $t\bar{t} \rightarrow e \nu \mu \nu~+$~2-jet category, with an additional
requirement that at least one jet must be tagged by the MV2c10
algorithm at the 85\% single-cut efficiency OP. The
LH method also exploits events with
exactly three jets, as well as events with same-flavour leptons in the 
final state. For events with same-flavour leptons in the final state,
additional requirements of \met$>$ 60~GeV and dilepton invariant mass
$50 < m_{\ell\ell} < 80$~GeV or
$m_{\ell\ell} > 100$~GeV
are applied to suppress the contamination from on-shell $Z$ boson
decays, multijet production and decays of
$\gamma^{\ast}, \Upsilon$ and $J/\psi$ particles. 
In the $e\mu$ + 2-jet channel of the LH (T\&P) method, a \ttbar\ purity of 82\% (90\%) is obtained, with sub-dominant contributions from single top, \zjets\ and diboson processes and events containing a misidentified lepton.  In the  $e\mu$ + 3-jet, \mbox{$\ell\ell$ + 2-jet} and \mbox{$\ell\ell$ + 3-jet} channels of the LH method, the \ttbar\ contribution is 88\%, 71\% and 79\%, respectively.
Figure~\ref{fig:PDFmodellingNoBDT} shows the jet \pt\
and $m_{\ell\ell}$ distributions for events passing the $e \nu \mu \nu~+
$~2-jet selection before any requirement on the MV2c10 output is applied.
Good agreement between the simulation and data is observed within the total statistical and systematic uncertainties.

\begin{table}
\renewcommand{\arraystretch}{1.1} 
\begin{centering}
  \begin{tabular}{l|c|c|c}
\hline
\hline
Selection Requirement & \multicolumn{2}{c|}{Likelihood Method} & T\&P Method\\
\hline
\hline
Leptons & \multicolumn{3}{c}{2 oppositely charged signal leptons ($e,\mu$)}\\
\hline
Jets & \multicolumn{2}{c|}{2 or 3 jets}  & 2 jets \\ \hline\hline
 & $e\mu$ & \eemumu\ & $e\mu$\\
\hline\hline
Region-specific cuts & & \met$>$ 60 GeV & At least 1 $b$-tagged jet\\
 & & $50 < m_{\ell\ell} < 80$~GeV $\vee$ $m_{\ell\ell} > 100$~GeV & (at 85\% efficiency OP)\\
\hline
BDT cut & \multicolumn{2}{c|}{${\cal{D}}_{bb}^\mathrm{LH}$ > 0.1} & ${\cal{D}}_{bb}^\mathrm{T\&P} > -0.02$\\
\hline 
\hline 
\end{tabular}
\caption{The analysis regions and associated event selection criteria for the LH and T\&P methods.  The variables ${\cal{D}}_{bb}^\mathrm{LH}$ and ${\cal{D}}_{bb}^\mathrm{T\&P}$ are BDT output discriminants trained to separate final states
with at least two \bjets\ from all other
events, in the LH and T\&P methods, respectively.  More details of the training and performance of the BDTs is presented in Section~\ref{sec:bdtraining}.}
\label{tab:Selections}
\end{centering}
\end{table}

For the purposes of the calibration methods, simulated events are categorised using generator level information according to the flavours of the selected reconstructed jets. For the two-jet selection, three 
possible jet flavour combinations are considered: $bb$, $bj$ and $jj$, where $b$
represents a $b$-jet, and $j$ is defined as a non $b$-jet.  In the selected events, the $c$-jet contribution is sufficiently small that it can be considered together with the light-flavour jet component. In the
same manner, for the three-jet selection, four possible jet flavour
combinations are considered: $bbb$, $bbj$, $bjj$ and $jjj$. The
dominant \ttbar\ process contributes mostly to the
$bb$ and $bj$ flavour combinations in the 2-jet channels, and to the $bbj$ and $bjj$ flavour combinations in the 3-jet channels. After adding all
background processes, especially from \zjets\
decays, the fraction of non-$b$-flavour jets increases. The overall
$b$-jet purity is calculated from simulation as the fraction of jets
in the sample that are labelled as $b$-jets. In the selected samples of $e\mu$ events,
it is found to be 72\% ({82\%}) in the two-jet case for the LH (T\&P) method,
while in the three-jet case it is 53\%. For the events with
\eemumu\ final states, the overall \bjet\ purity is 62\% and
48\% in the two-jet and three-jet cases, respectively. 
As the number of background events containing at least one misidentified lepton is estimated from data, and therefore the jet flavour composition is unknown, it is assumed that only non-$b$-jets
are produced in these events. 
A cross-check was performed assuming that only $b$-jets
 are produced in these events, and found to have a negligible effect on the calibration results.

\subsection{Multivariate event discriminant}
\label{sec:bdtraining}

In order to further enhance the $b$-jet purity of the selected samples
in both methods, boosted decision trees (BDT) were trained using
simulated events to separate final states
with at least two \bjets, defined as the signal, from all other
events, classified as a
background. Each of the input variables is designed to select events
with at least two \bjets\ based upon the topology and kinematics of the event, rather than exploiting any flavour-tagging-related properties of the jets, to ensure minimal bias in the MV2c10 discriminant.
A dedicated optimisation was performed for each method, leading to a different choice of input variables, as shown in Table~\ref{tab:BDTVariables}.  In both the LH and T\&P methods, the BDTs are trained using the Toolkit for Multivariate Data Analysis, TMVA~\cite{TMVA}.

In the LH method, sample-selection BDTs are trained separately for the two- and
three-jet categories. The training samples include not only the
nominal \ttbar\ sample, but also the alternative \ttbar\ samples
used for evaluating systematic
uncertainties. The inclusion of the alternative \ttbar\ samples provides a larger training sample and allows the BDT to learn the topologies of events generated using
alternative MC generators, helping to minimise uncertainties in the calibration due to the choice of generator.

\begin{figure}[h]
\begin{center}
\subfloat[][]{
\includegraphics[width=0.495\textwidth]{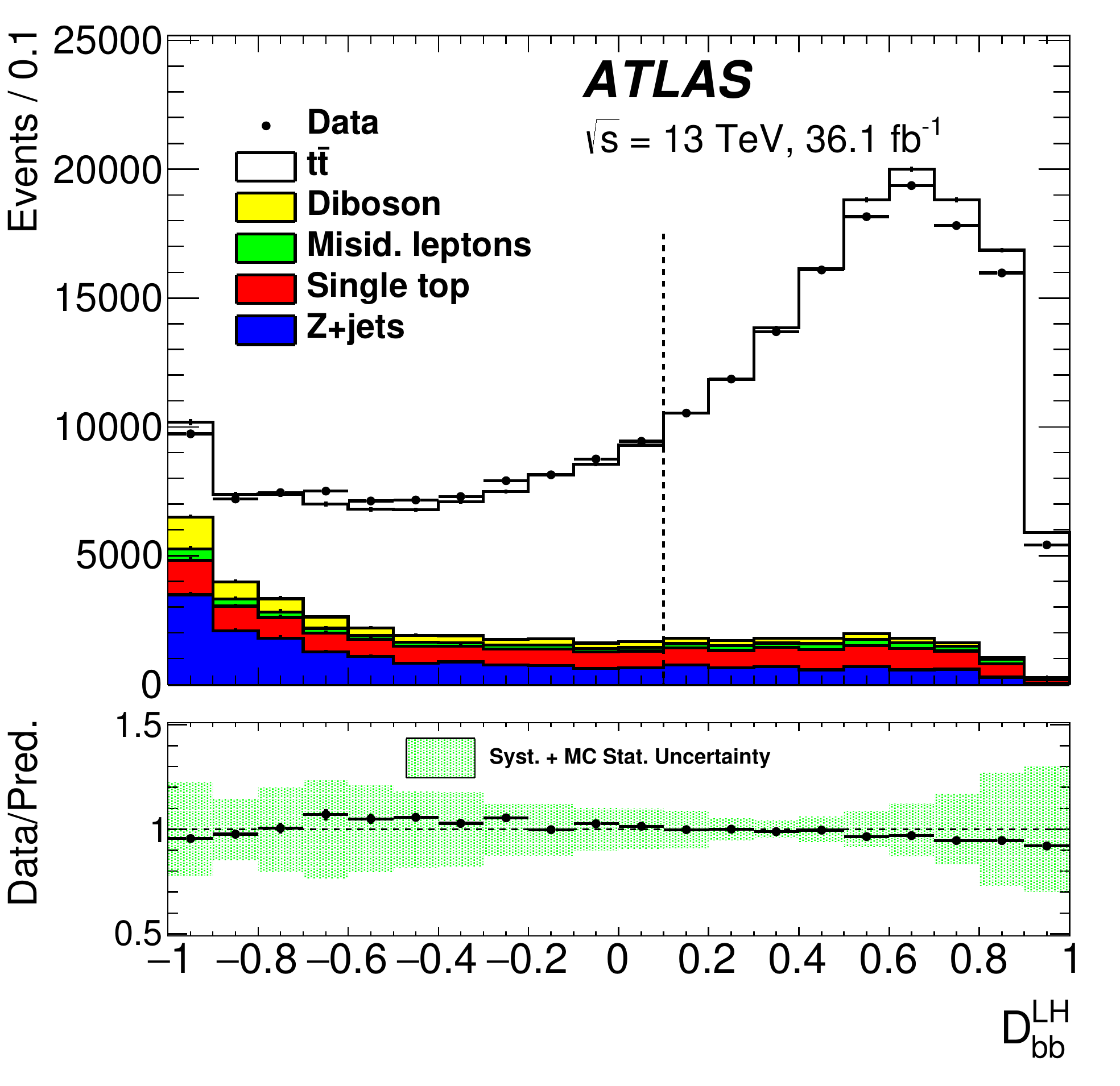}
\label{fig:BDTLH}
}
\subfloat[][]{ 
\includegraphics[width=0.495\textwidth]{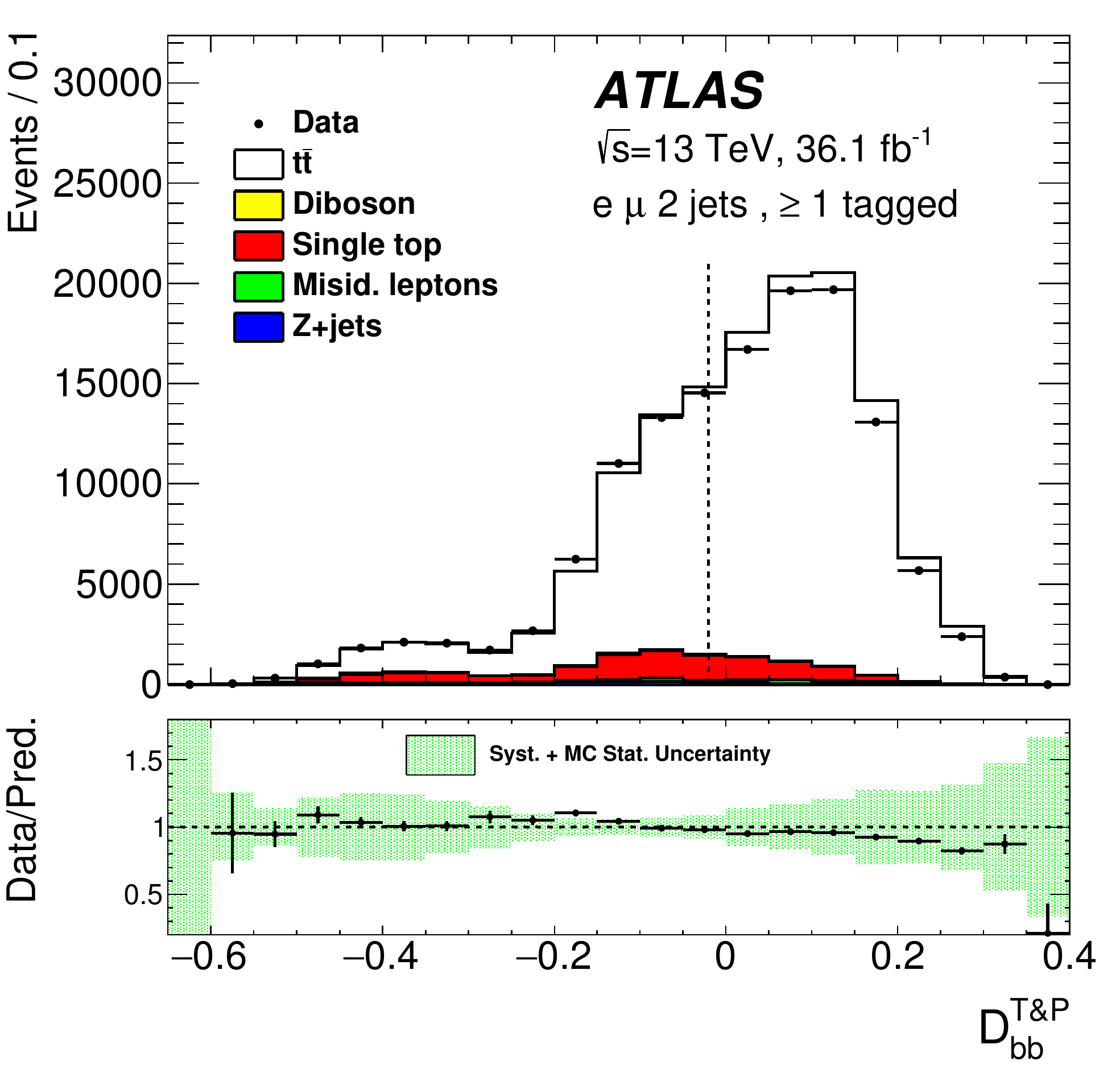}
\label{fig:BDTTP}
}
\caption{The sample-selection BDT output distribution for the data
  (points) and simulated samples (stacked histograms) in (a) the
  combined $ee/e\mu/\mu\mu+2/3$-jets sample used in the LH method and (b)
  for the $e\mu+2$-jets sample used in the T\&P method.  The
  simulated samples in the LH method are normalised to agree with a fit to
  the data as described in Section~\ref{subsec:PDFmethod}, while in the T\&P method, the normalisation is taken from the theoretical predictions. The cut applied to the ${\cal{D}}_{bb}^{\mathrm{LH}}$  (${\cal{D}}_{bb}^\mathrm{T\&P}$) discriminant in the LH (T\&P) method is indicated by a vertical dotted line.  The bottom
  panels show the ratio of the data to the simulated samples, with the
 shaded uncertainty bands corresponding to the total MC statistical uncertainty and
  systematic uncertainty.}
\label{fig:BDT}
\end{center}
\end{figure}

For both the LH and T\&P methods, all input variables, as well as ${\cal{D}}_{bb}^{\mathrm{LH}}$ and ${\cal{D}}_{bb}^\mathrm{T\&P}$, are well modelled in simulation.
The modelling of ${\cal{D}}_{bb}^\mathrm{LH}$ is shown in
Figure~\ref{fig:BDTLH}, combining all four event categories used in the LH method. Good agreement between the simulation and data is observed in all four channels.  Only events with ${\cal{D}}_{bb}^\mathrm{LH}>0.1$ are considered in the measurement. This
threshold is found to be optimal, as it minimises the
total uncertainty in the measurement of the calibration scale factor.

The modelling of
${\cal{D}}_{bb}^\mathrm{T\&P}$ is shown in Figure~\ref{fig:BDTTP}. The
selection requirement on the BDT output was optimised such that it minimises the
uncertainty associated with the data-to-simulation scale factors, and is found to be ${\cal{D}}_{bb}^\mathrm{T\&P}>-0.02$.

\begin{figure}[h]
\begin{center}
\subfloat[][]{
  \includegraphics[width=0.50\textwidth]{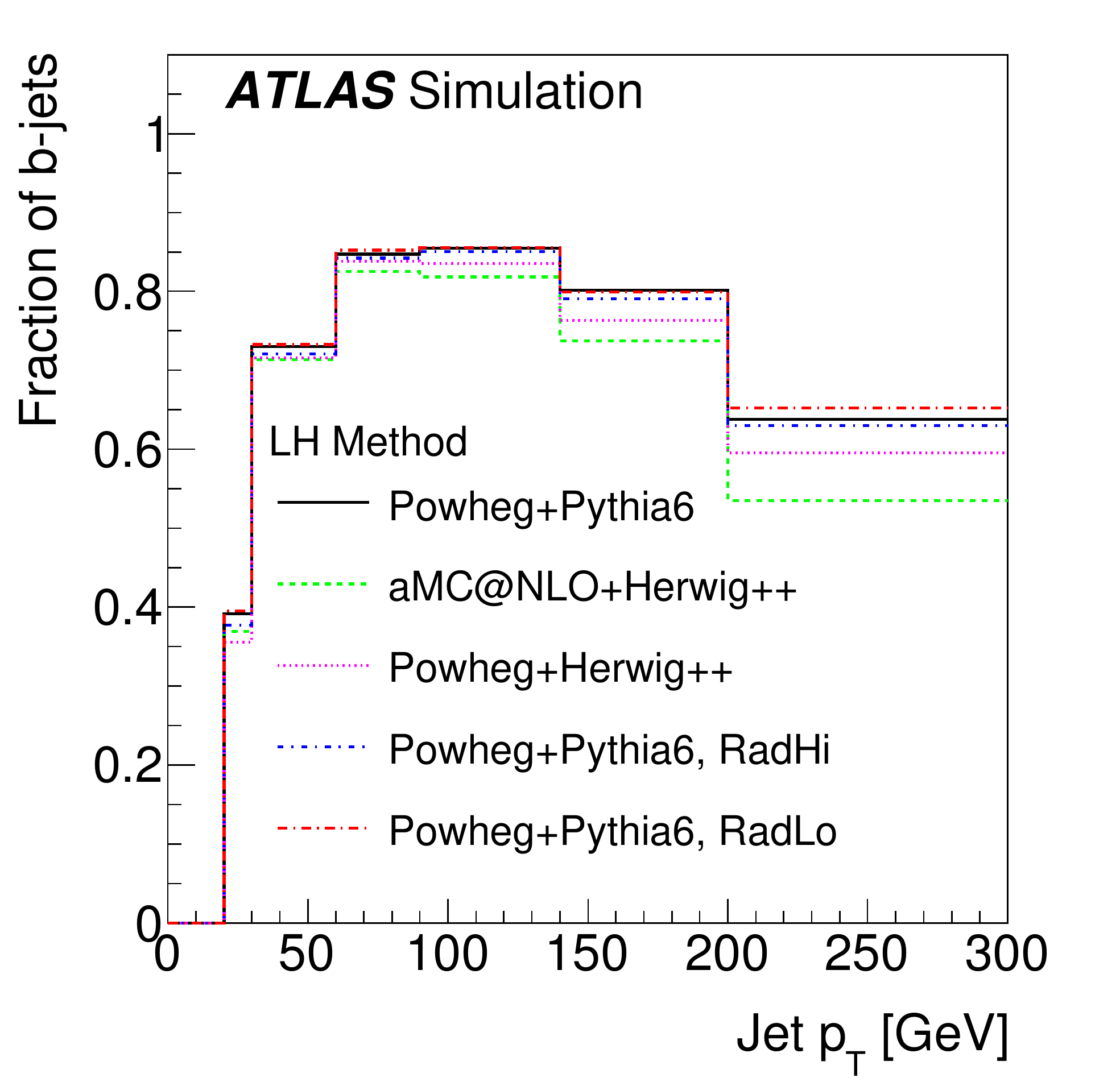}\label{fig:Purity_a}}
\subfloat[][]{
  \includegraphics[width=0.50\textwidth]{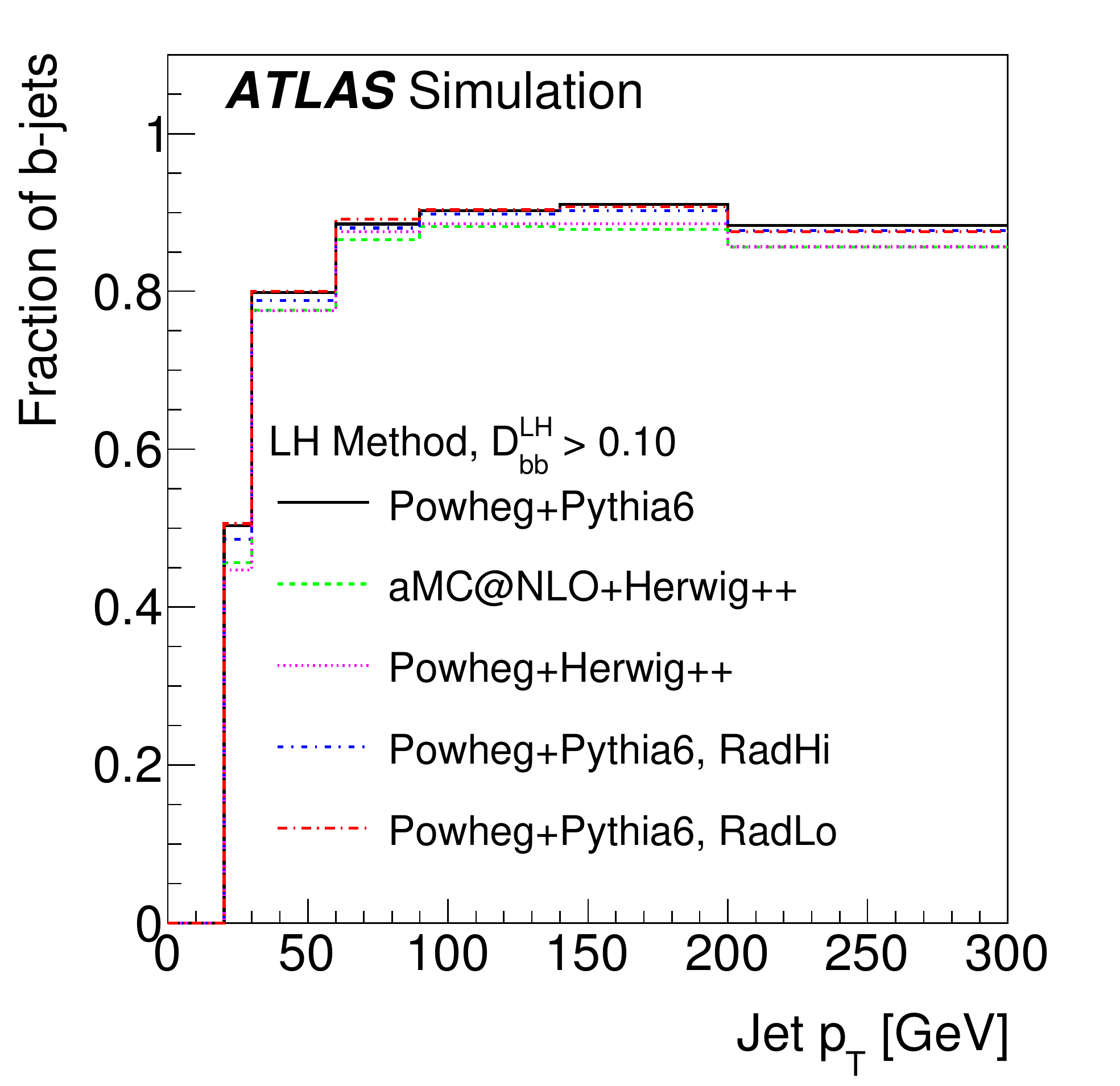}\label{fig:Purity_b}} \\
\subfloat[][]{
  \includegraphics[width=0.50\textwidth]{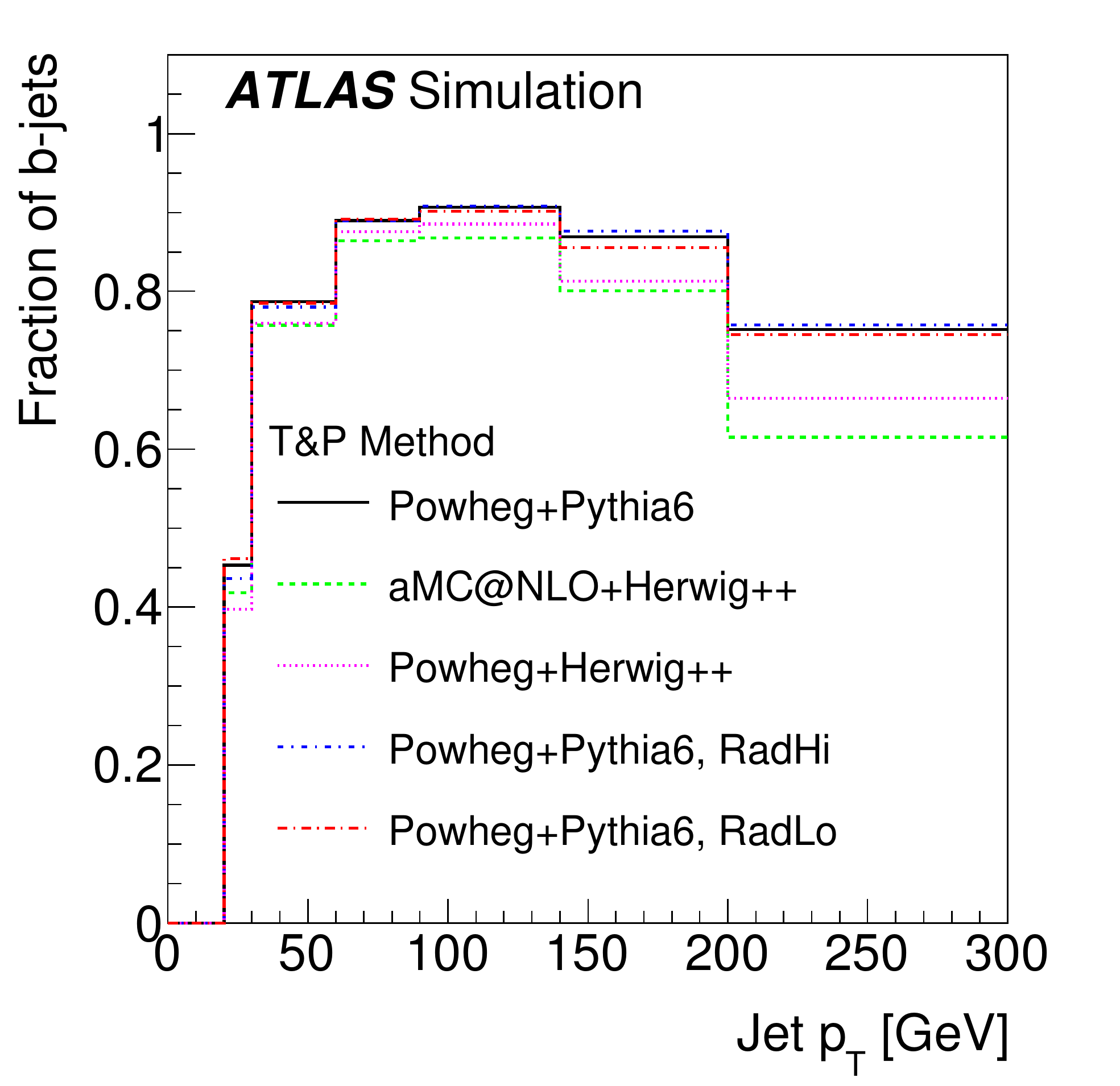}\label{fig:Purity_a_TP}}
\subfloat[][]{
  \includegraphics[width=0.50\textwidth]{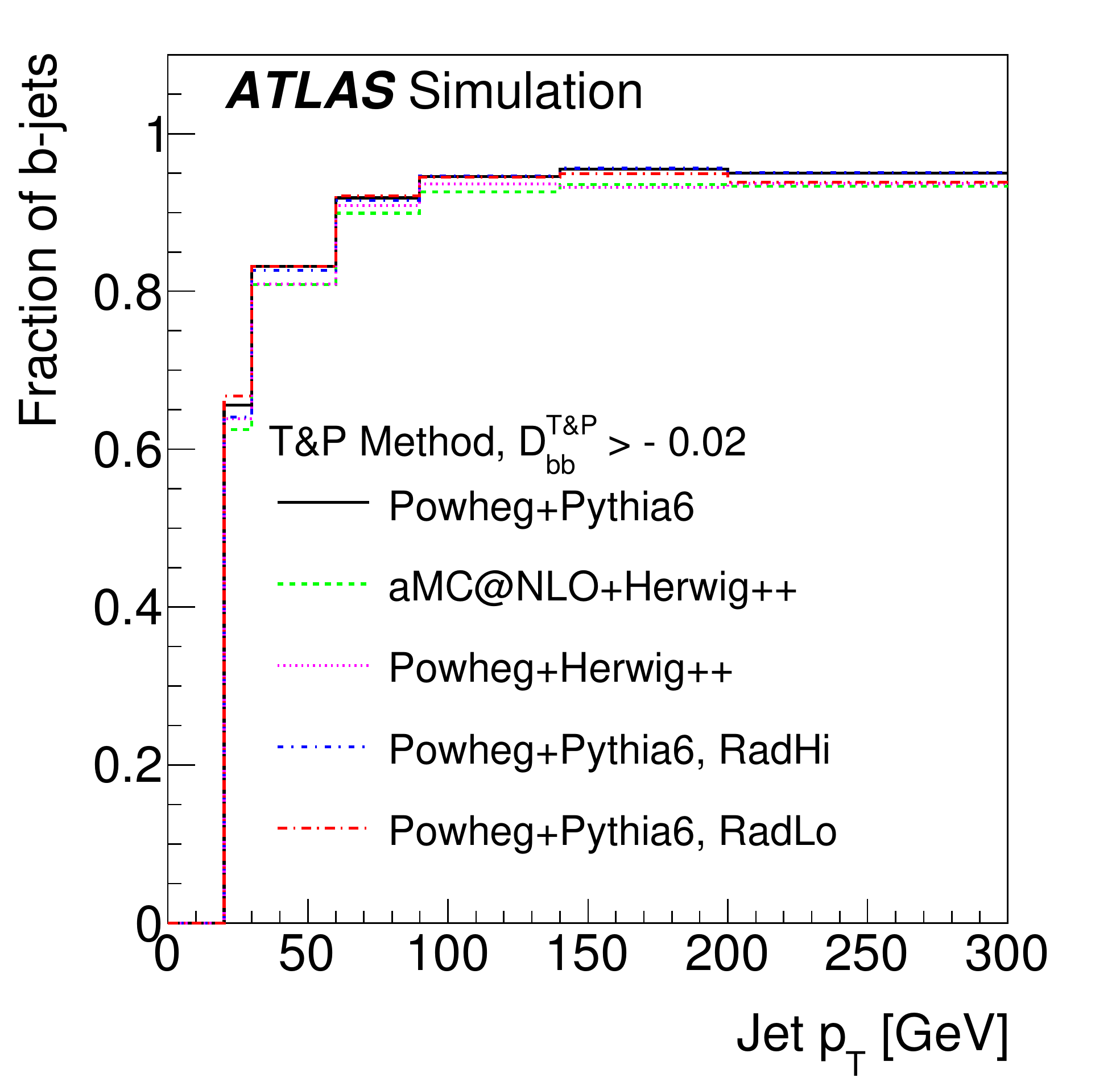}\label{fig:Purity_b_TP}}
\caption{The fraction of $b$-jets in the selected sample as a function
  of the jet \pt, (a) before and (b) after the requirement on ${\cal{D}}_{bb}^\mathrm{LH}$, for the
  nominal 
   and alternative
    \ttbar\ generators in the
  $e\mu+2$-jet category of the LH method.  The fraction of $b$-jets in the selected sample as a function
  of the jet \pt,  (c) before and (d) after the requirement on ${\cal{D}}_{bb}^\mathrm{T\&P}$, for the
  nominal
   and alternative 
   \ttbar\ generators in the
  $e\mu+2$-jet category of the T\&P method.}
  \label{fig:PDFmodelling}
\end{center}
\end{figure}

Figure~\ref{fig:PDFmodelling}
shows the fraction of $b$-jets in
bins of jet \pt\ both before and after applying a requirement on
${\cal{D}}_{bb}^\mathrm{LH}$ (${\cal{D}}_{bb}^\mathrm{T\&P}$) for the nominal and alternative \ttbar\ generators in
the $e\mu+2$-jet category of the LH (T\&P) method.

\begin{table}[h]
\renewcommand{\arraystretch}{1.1} 
\begin{centering}
\begin{tabular}{l|c|c|c}
\hline \hline
Variable & Definition & T\&P & LH \\ \hline \hline
$\ell_{1}$  \pt\ & Leading lepton \pt\ & $\times$ &  \\
 Jet$_{1}$ \pt\ & Leading jet \pt\ & $\times$ & $\times$ \\
Jet$_{2}$ \pt\ & Sub-leading jet \pt\ & $\times$ & $\times$ \\
Jet$_{3}$ \pt\ & Third-leading jet \pt\ (events with 3 jets only) &  & $\times$ \\
\met\ & Missing transverse momentum & $\times$ &  \\
$n_{\text{Fjets}}$ & Number of jets with $2.5 < |\eta| < 4.5$ & $\times$ &  \\
$\Delta \phi ({j}_{1}, {j}_{2})$ & $\Delta\phi$ of leading 2 jets & & $\times$ \\
$\min\Delta R ({j}, {j})$ & Minimum $\Delta R$ of all jet combinations  &  & $\times$ \\
Imbalance & (jet$_{1}(p_\mathrm{T}) - $jet$_{2}(p_\mathrm{T})) / ($jet$_{1}(p_\mathrm{T}) + $jet$_{2}(p_\mathrm{T}))$  &  & $\times$ \\
$m_\mathrm{min}^\mathrm{avg}(\ell j)$ & min(($m(\ell_{1},$jet$_{i})+m(\ell_{2},$jet$_{k})$)/2), & $\times$ & $\times$ \\
 & $i,k=1,2$ $(1,2,3)$ for events with 2~(3)~jets & &\\
$\mathrm{min} \Delta R(\ell_1,j)$ & Minimum $\Delta R$ separation of leading lepton from all jets & $\times$ & $\times$\\
$\mathrm{min} \Delta R(\ell_2,j)$ & Minimum $\Delta R$ separation of subleading lepton from all jets & & $\times$\\
\hline 
\hline
\end{tabular}
\caption{Input variables used in the T\&P and LH method BDT
  algorithms. The ``$\times$'' symbol in the T\&P and LH columns
  indicates the BDT in which each variable is used.  Quantities involving jets labelled `$j$' correspond to central jets, with $|\eta| < 2.5$.}
\label{tab:BDTVariables}
\end{centering}
\end{table}

\clearpage


\section{Calibration methods}
\label{sec:methods}

\subsection{Tag-and-probe method}
\label{subsec:TPmethod}
The sample used by the T\&P method is 90\% pure in \ttbar\ events, providing a high-purity sample of
$b$-jets.
The \beff\ measurement is performed on a set of \textit{probe} jets, where a jet is considered a \textit{probe} jet if the other jet is $b$-tagged at the 85\% efficiency OP.

\begin{figure}
\begin{center}   
\subfloat[][]{
  \includegraphics[width=0.50\textwidth]{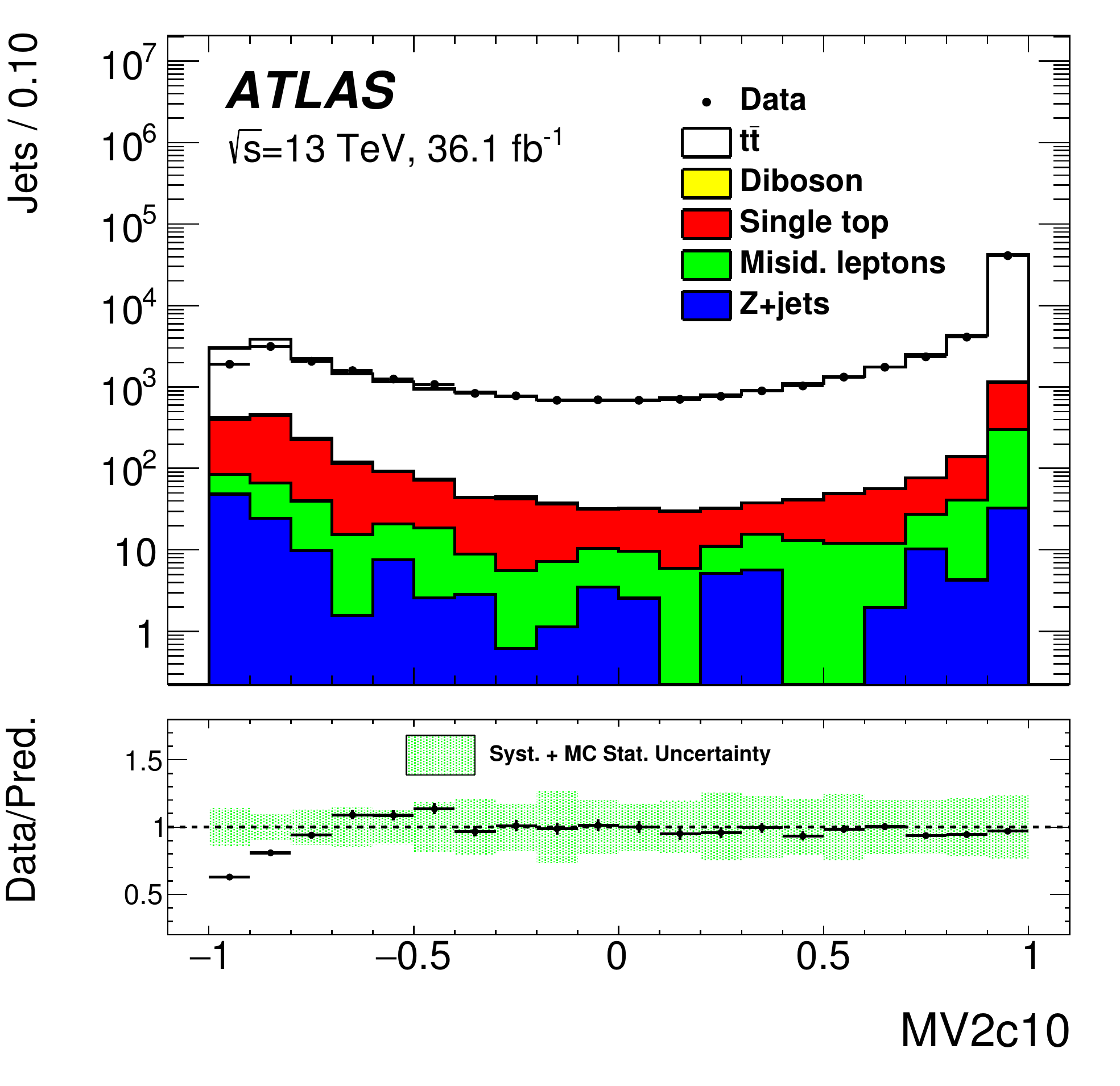}\label{fig:mv2c10_comp}}
\subfloat[][]{
  \includegraphics[width=0.50\textwidth]{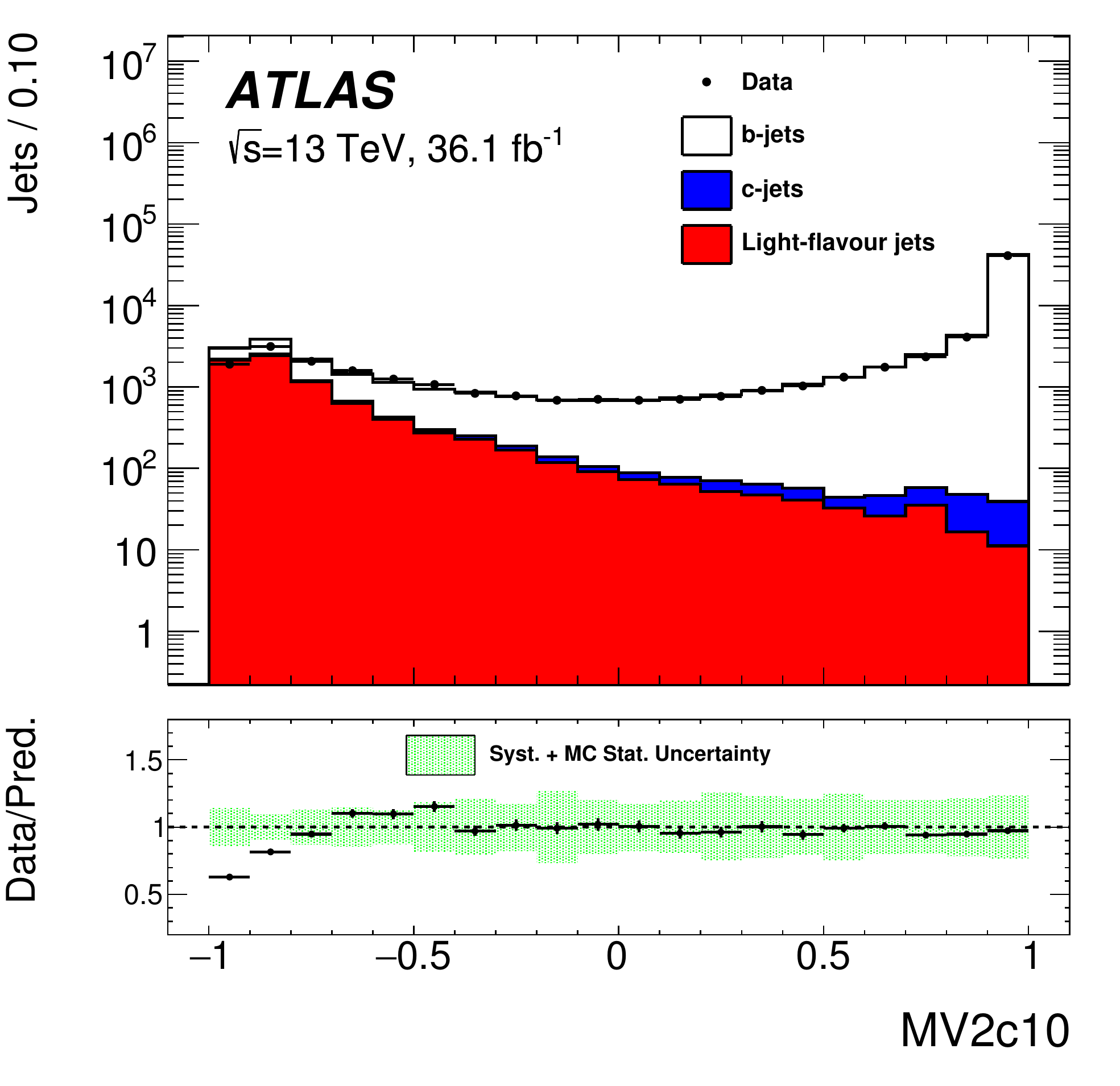}\label{fig:mv2c10_flav}} \\
\caption{The MV2c10 distribution of the probe jets used for the
  calibration in the T\&P method, (a) broken down by process, and (b) probe jet flavour.
    }
\label{fig:probeMV2c10}
\end{center}
\end{figure}

The MV2c10 distributions for probe jets in data and simulation are presented in Figure ~\ref{fig:probeMV2c10}, broken down by both process and probe jet flavour.  Good agreement between simulation and data is observed in the $b$-jet dominated MV2c10 output region.  In the light-flavour jet dominated MV2c10 output region, some disagreement between simulation and data is observed, however, this is accounted for by the light-flavour jet SFs and associated uncertainties~\cite{PERF-2012-04}.  The $b$-tagging efficiency is measured by first determining the fraction of the probe jets that satisfy a given $b$-tagging criterion, \mbox{$f_{\mathrm{tagged}}=N^{\mathrm{pass}}/N$}, where $N$ is the total number of probe jets and $N^{\mathrm{pass}}$ is the number of probe jets satisfying the criterion. This fraction is measured in 
data by subtracting the contribution from
non-\ttbar\ processes, as predicted by simulation
 \begin{equation*}
f_{\mathrm{tagged}}= \frac{N_{\mathrm{data}}^{\mathrm{pass}}-N_{\text{non-}t\bar{t},\mathrm{sim}}^{\mathrm{pass}}}{N_{\mathrm{data}}-N_{\text{non-}t\bar{t},\mathrm{sim}}},
\label{efficTP}
\end{equation*}
\noindent where $N_{\mathrm{data}}$ is the number of probe jets in data and
$N_{\text{non-}t\bar{t},\mathrm{sim}}$ is the number of probe jets from
non-\ttbar\ events predicted by simulation.  It should be noted that non-\ttbar\ processes can contribute \bjets\ as well as light-flavoured jets.

The determination of the \beff\ relies on the assumption that the fraction of probe jets containing a tagged jet in data, $f_{\mathrm{tagged}}$,
is given by
\begin{equation*}
f_\mathrm{tagged} = f_{b} \varepsilon_{b} + (1-f_{b}) \varepsilon_{j}, 
\label{simpleTagFrac}
\end{equation*}
\noindent where $ f_{b}$ and $
(1-f_{b})$ are the fractions of $b$-jets
and non-$b$-jets in \ttbar\ events, and $\varepsilon_{b}$ and $\varepsilon_{j}$ are
the $b$-jet and non-$b$-jet tagging
efficiencies. The $b$-jet tagging efficiency can be determined by
measuring $f_{\mathrm{tagged}}$ in data, and estimating the other parameters from
simulation. In this approach, jets are considered as uncorrelated
objects within an event. The $b$-tagging efficiency is extracted
from $f_\mathrm{tagged}$ in bins of the jet \pT
\begin{equation*}
\varepsilon_{b} = \frac{f_\mathrm{tagged} - (1-f_{b}) \varepsilon_{j}}{f_{b}}.
\label{simpleTagEffic}
\end{equation*}
The main sources of systematic uncertainty arise from the
determination of $f_{b}$ and $ \varepsilon_{j}$, as both of these parameters
are taken from simulation.

\subsection{Combinatorial likelihood method}
\label{subsec:PDFmethod}
 The LH method is performed separately for the $e\mu$ and combined
\eemumu\ final states in the two- and three-jet bin categories due to differences in the
background composition. The yields of \ttbar\ and
\zjets\ events are normalised using
dedicated control samples in data. The selection criteria applied to define
these control regions are given in Table~\ref{tab:LM_CRs}. Normalisation factors are determined simultaneously with the
maximum-likelihood fit to the observed number of events in all
regions, and used to scale the overall normalisation of the \ttbar\ and \zjets\ processes. The values of the fitted normalisation factors, along with their
associated statistical uncertainties, are also presented
in Table~\ref{tab:LM_CRs}.

\begin{table}
\begin{centering}
\begin{tabular}{l|c|c}
\hline \hline
Control sample & $e\mu$ & \eemumu\ \\ \hline
\ttbar\ & $e\mu$ selections & \eemumu\ selections before ${\cal{D}}_{bb}^\mathrm{LH}$ requirement \\
&before ${\cal{D}}_{bb}^\mathrm{LH}$ requirement &\met$>60$~GeV \\
&  & $50 < m_{\ell\ell} < 80$~GeV, or $m_{\ell\ell}>100$~GeV \\
NF (2-jet) & 0.924 $\pm$ 0.004 & 0.906 $\pm$ 0.007 \\
NF (3-jet) & 1.010 $\pm$ 0.004 & 0.986 $\pm$ 0.008 \\
\hline
\zjets\ & Same lepton flavour,& Same lepton flavour,\\
& $80 < m_{\ell\ell} < 100$~GeV &\met$>60$~GeV, $80 < m_{\ell\ell} <
                                  100$~GeV \\

NF (2-jet) & 1.104 $\pm$ 0.001 & 1.096 $\pm$ 0.005 \\
NF (3-jet) & 1.258 $\pm$ 0.002 & 1.281 $\pm$ 0.008 \\
\hline
\end{tabular}
\caption{Definition of the control samples in data used for
 the determination of the normalisation factors for $t\bar{t}$ and
\zjets\ processes in the LH method. The obtained normalisation factors
(NFs) with their
associated statistical uncertainties in each category are also presented.}
\label{tab:LM_CRs}
\end{centering}
\end{table}

Unlike the T\&P method, the LH method exploits the per-event jet
correlations. For example, in the case of events with two jets, the fraction of events with one $b$-tagged jet, $f_{\text{1-tag}}$, or two $b$-tagged jets, $f_{\text{2-tag}}$, can be measured in data using 
\begin{align*}
\label{noSimpleTag}
  f_{\text{1-tag}} & =  2 f_{bb} \varepsilon_{b} \left( 1-\varepsilon_{b} \right) + 
  f_{bj} \left[ \varepsilon_{j} \left( 1 - \varepsilon_{b} \right) + 
    \left( 1 - \varepsilon_{j} \right) \varepsilon_{b} \right] +
  \left(1 - f_{bb} - f_{bj}\right) 2 \varepsilon_{j} 
  \left( 1 - \varepsilon_{j} \right)\\
\nonumber 
f_{\text{2-tag}} & =  f_{bb}\varepsilon^2_{b} + f_{bj} \varepsilon_{j} \varepsilon_{b} + 
  \left(1 - f_{bb} - f_{bj}\right) \varepsilon^2_{j},
\end{align*}
where $f_{bj}$ and $f_{bb}$ are the fraction of events with one $b$-jet and two $b$-jets, respectively, and $\varepsilon_{b}$ ($\varepsilon_{j}$) is the $b$- (light-flavour) jet tagging efficiency.  Using these equations, $\varepsilon_{b}$ is determined by measuring $f_{1-\mathrm{tags}}$ and $f_{\text{2-tag}}$ from data, with $f_{bj}$, $f_{bb}$ and $\varepsilon_{j}$ determined from MC.
In this way, the correlation of the jet flavour information in the 1-tag and 2-tag regions is added, resulting in a more precise efficiency measurement. A measurement in $N$ kinematic bins
results in $2 \times N^{2}$ coupled
equations. It is possible to solve such a system of non-linear equations, but in practice it is much simpler to model the same system by using a more flexible and powerful likelihood function and solve the system numerically by maximising the likelihood.

Using the probability density functions, $\mathcal{P}$, the per-event
likelihood term for the two jets in the event to have transverse momenta $p_\mathrm{T,1}$ and
$p_\mathrm{T,2}$ and MV2c10 weight outputs $w_{1}$ and $w_{2}$  is defined as
\begin{eqnarray*}
\label{likelihood}
  {\cal L}_{\mathrm{event}}\left(p_{\mathrm{T},1},p_{\mathrm{T},2},w_{1},w_{2}\right) 
  = &[& f_{bb} {\mathcal{P}}_{bb}\left(p_{\mathrm{T},1},p_{\mathrm{T},2}\right) 
    {\mathcal{P}}_{b}\left(w_{1}\vert p_{\mathrm{T},1}\right) 
    {\mathcal{P}}_{b}\left(w_{2}\vert p_{\mathrm{T},2}\right) \\
    & + &   f_{bj} {\mathcal{P}}_{bj}\left(p_{\mathrm{T},1},p_{\mathrm{T},2}\right) 
    {\mathcal{P}}_{b}\left(w_{1}\vert p_{\mathrm{T},1}\right) 
    {\mathcal{P}}_{j}\left(w_{2}\vert p_{\mathrm{T},2}\right)  \nonumber \\
    & + &   f_{jj} {\mathcal{P}}_{jj}\left(p_{\mathrm{T},1},p_{\mathrm{T},2}\right) 
    {\mathcal{P}}_{j}\left(w_{1}\vert p_{\mathrm{T},1}\right) 
    {\mathcal{P}}_{j}\left(w_{2}\vert p_{\mathrm{T},2}\right) \nonumber \\ 
    & + &   1  \leftrightarrow 2 ~ ]/2~, \nonumber
\end{eqnarray*}
where ${\mathcal{P}}_{f_{1}f_{2}}(p_\mathrm{T,1},p_\mathrm{T,2})$ is a two-dimensional
probability density function for jets
of flavour $f_{1}$ and $f_{2}$ to have transverse momenta $p_\mathrm{T,1}$ and
$p_\mathrm{T,2}$, and ${\mathcal{P}}_{f}(w,p_\mathrm{T})$ is a probability density function of the $b$-jet tagging
weight for a jet of flavour $f$ at a given \pt. The factors $f_{bb}$, $f_{bj}$,
and $f_{jj} = 1 - f_{bb} - f_{bj}$ are the overall flavour fractions in the two-jet
case. All probability density functions are determined from simulation, except the one for the $b$-jet
weight, which contains the information to be extracted from
data.

The $b$-jet tagging efficiency corresponding to the MV2c10 weight cut of $w_\mathrm{cut}$ is given by

\begin{equation*}
\varepsilon_{b} (p_\mathrm{T})=\int_{w_\mathrm{cut}}^{\infty} {\text{d}w'{\mathcal{P}}_{b}(w',p_\mathrm{T})}.
\label{efficPDF}
\end{equation*}

For the extraction of the \bjet\ tagging efficiency for a single OP,
 $\mathcal{P}_{b}(w',p_\mathrm{T})$ corresponds
to a histogram with two $w'$ bins for each jet \pT\ bin. The bin above $w_\mathrm{cut}$ corresponds to the \beff.

For events containing three jets, the likelihood is constructed in a way
analogous to the two-jet case, resulting in six equivalent likelihood
terms instead of four.

A closure test was performed by applying the full method to the
simulated events. 
This sample of simulated events is normalised to an integrated luminosity of~36.1~fb$^{-1}$, and
is treated
 as ``pseudo data'' with the expected number of events in each bin taken as the mean of a Poisson distribution to estimate the statistical uncertainty.
The resulting scale factors are close to unity within the statistical uncertainty of the pseudo data sample for all bins except the lowest jet-\pt\ bin in the 3-jet sample, verifying that the method has no significant bias. An additional 3\% uncertainty is added to cover the observed non-closure in the lowest \pt\ bin for the three-jet sample.

\section{Systematic uncertainties}
\label{sec:systematics}

 Three categories of systematic uncertainty are considered in the measurements presented. First, \textit{MC generator modelling uncertainties} that affect the modelling of kinematic distributions and the jet flavour composition in simulated events. Second, \textit{normalisation uncertainties} that account for uncertainties in the cross-section of simulated samples. Third, \textit{experimental uncertainties}, which are related to detector effects and the reconstruction of the physics objects in the simulated samples. 

Uncertainties from the MC generator modelling are evaluated in the simulated \ttbar, single-top, and \zjets\ samples, by comparing the nominal samples to ones created with alternative generators and settings.  The alternative generators induce changes in the event kinematics and flavour composition, thereby affecting the extracted scale factors. The difference between the scale factors is taken as the systematic uncertainty.  Uncertainties are estimated for the predictions from the nominal \ttbar\ sample (\textsc{Powheg+Pythia6}), by altering the choice of parton shower and hadronisation generator (\textsc{Powheg+Herwig++}) or by altering the matrix element generator (\textsc{MG5}\_a\textsc{MC@NLO+Herwig++}). Changing the settings of the nominal generator to increase or decrease the amount of parton radiation and varying the choice of parton distribution function (PDF) set (\textsc{MG5}\_a\textsc{MC@NLO+Herwig++} with \textsc{CT10 PDFs} or \textsc{PDF4LHC15\_NLO PDFs}) gives additional sources of uncertainty.
Further uncertainties due to the observed mismodelling of the top quark and \ttbar-system \pt\ are evaluated by taking the difference between the default \ttbar\ prediction and a sample in which the top quark and \ttbar-system \pt\ distributions are reweighted to match predictions at NNLO accuracy in QCD~\cite{NNLOpaper,Czakon:2016dgf}. For $Wt$ single-top production, uncertainties are estimated by varying the parton shower and hadronisation modelling (\textsc{Powheg+Herwig++}) and varying settings of the nominal generator to increase or decrease the amount of parton radiation. The uncertainty in the treatment of the interference between $Wt$ and \ttbar\ is assessed by replacing the nominal diagram removal (DR) scheme with a diagram subtraction (DS) scheme~\cite{Frixione:2008yi} in \textsc{Powheg+Pythia6}. An additional 6\% uncertainty is applied to the normalisation of the single-top samples to account for the uncertainty in the predicted cross-section~\cite{Kidonakis:2010ux}. For the \zjets\ process, the nominal samples (\textsc{MG5}\_a\textsc{MC@NLO}) are compared to the alternative \textsc{Powheg+Pythia8} sample. In addition an uncertainty is estimated for the modelling of the jet \pt\ spectrum in the \zjets\ events with the same-flavour leptons in the final state by reweighting the spectrum to match the data in the \zjets\ control sample. 

To account for the extrapolation of the \zjets\ normalisation from the control sample to the sample used in the $b$-tagging efficiency measurement, a 20\% uncertainty in the \zjets\ estimate is applied.  The size of this uncertainty is determined by comparing the data to MC simulation in the relevant kinematic distributions. 
An additional 50\% uncertainty is applied to the events with at least one $b$- or $c$-jet, as observed in $Z+b$ measurements~\cite{Aad:2011jn}. Due to the small contribution from the diboson backgrounds, only a normalisation uncertainty is assigned to this sample. This uncertainty is assumed to be 50\% in the two-jet channel, and 70\% in the three-jet channel, as determined from MC studies. Likewise for the backgrounds with misidentified leptons, only a normalisation uncertainty of 50\% is considered.  Identical normalisation uncertainties are applied in the LH and T\&P methods.

In addition, 
in the T\&P method, which has a tighter event selection than the LH method, the uncertainties arising from the limited size of the simulated samples have a non-negligible effect on the order of 1\% on the total scale factor uncertainty. They are evaluated using 10,000 pseudo experiments. In each pseudo experiment the efficiency and the data-to-simulation scale factor in each jet \pt\ bin is computed. The standard deviation of the scale factor in all of the pseudo experiments is taken as the systematic uncertainty due to limited MC sample size.  The impact of MC statistical uncertainties is significantly lower in the more inclusive LH method, and therefore the impact of MC statistical uncertainties in the likelihood model itself is not considered.

The experimental uncertainties include those related to the reconstruction of electrons, muons, jets and \met, uncertainties in the mis-tagging of $c$- and light-flavour jets as $b$-jets, uncertainties in the modelling of pile-up and in the integrated luminosity.
For both the electrons~\cite{ATLAS-CONF-2016-024} and muons~\cite{PERF-2015-10}, uncertainties are estimated for the energy scale and resolution, as well as the reconstruction, identification, and trigger efficiencies using 13~TeV data.
The uncertainties in the jet energy scale and resolution are evaluated using 13~TeV data~\cite{Aaboud:2017jcu}, and so is an uncertainty in the efficiency of the JVT selection~\cite{PERF-2014-03}. The uncertainties in the energy scale and resolution of the jets and leptons are propagated to the calculation of the \met, which also has additional dedicated uncertainties from the momentum scale, resolution and efficiency of the tracks not associated with any of the reconstructed objects, along with the modelling of the underlying event. The predicted rate of mistakenly tagging non-$b$-jets is corrected using data-to-simulation efficiency scale factors measured separately for $c$-jets and light-flavour jets~\cite{PERF-2012-04}. The uncertainty in this prediction is estimated by varying these scale factors within their associated uncertainties.  For Run 2 data, the $c$-jet efficiency  scale factor uncertainty varies from $\sim 15\%$ for a jet \pt\ of 100~GeV, to $\sim 30\%$ for a jet \pt\ of 300~GeV.  For light-flavour jets, the uncertainty varies from $\sim 40\%$ for a jet \pt\ of 100~GeV, to $\sim 30\%$ for a jet \pt\ of 300~GeV.
The uncertainty due to the reweighting of the distribution of the expected average number of interactions per bunch crossing, $\langle \mu \rangle$,
from the simulation to the one measured in data is estimated by varying the nominal reweighting scale factor by the size of the nominal correction. The uncertainty in the combined 2015 and 2016 integrated luminosity is 3.2\%. It is derived, following a methodology similar to that detailed in Ref.~\cite{Aaboud:2016hhf}, from a preliminary calibration of the luminosity scale using $x$--$y$ beam-separation scans performed in August 2015 and May 2016. 

The effect of each source of systematic uncertainty on the $b$-jet tagging efficiency data-to-simulation scale factors is computed by replacing the nominal simulated sample with the sample affected by the systematic variation, and rerunning the fit to data. The uncertainty is taken as a difference relative to the scale factor measured in the nominal case.  When combining all four channels in the LH method, all single systematic variations are treated as fully correlated, except for the background modelling uncertainties, for which a 50\% correlation is assumed.  This partial correlation is applied, as each modelling variation is expected to account for more than one effect.


\section{Results}
\label{sec:results}

\begin{figure}
\begin{center}
\subfloat[][]{
\includegraphics[width=0.49\textwidth]{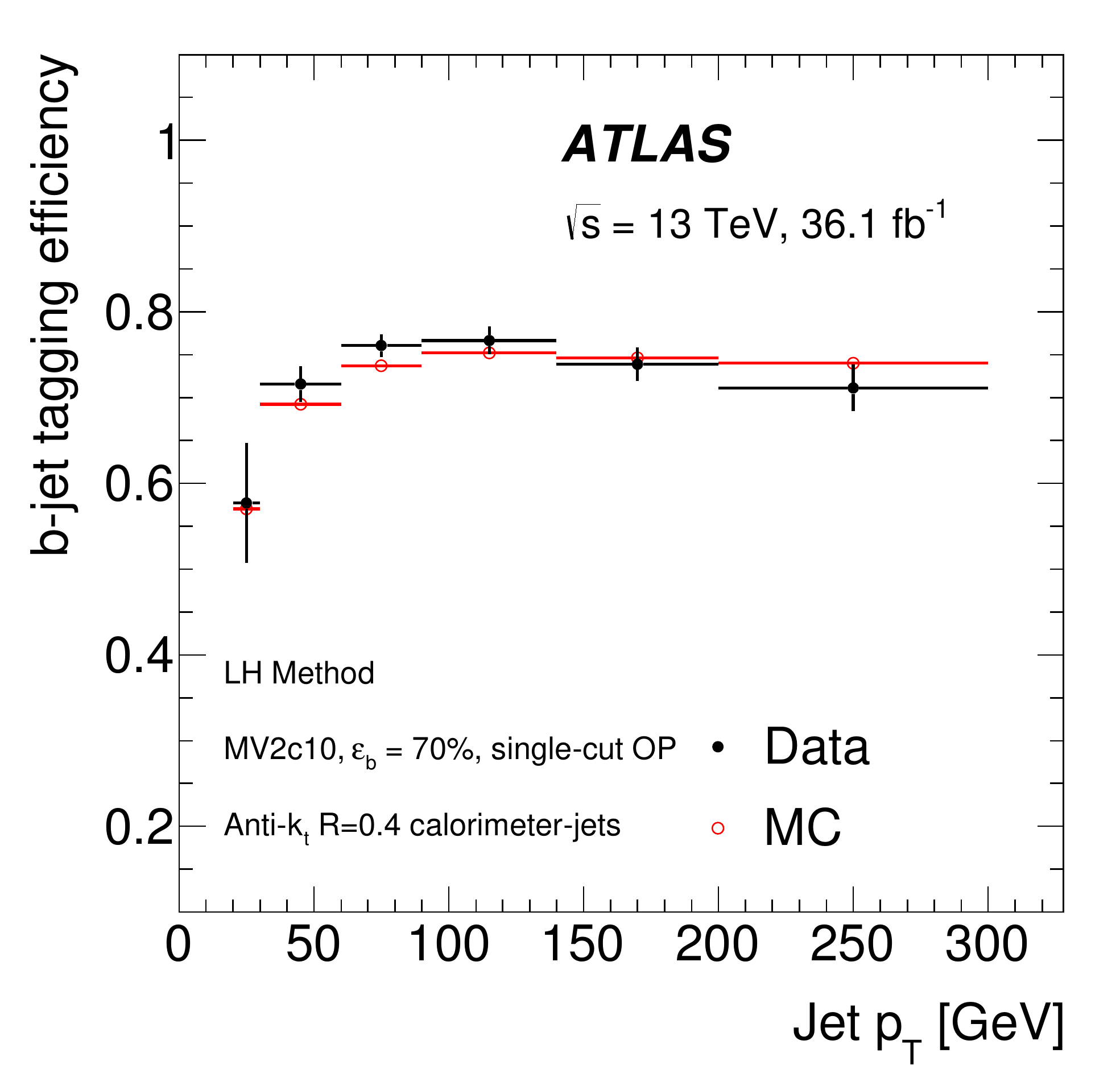}
}
\subfloat[][]{
\includegraphics[width=0.49\textwidth]{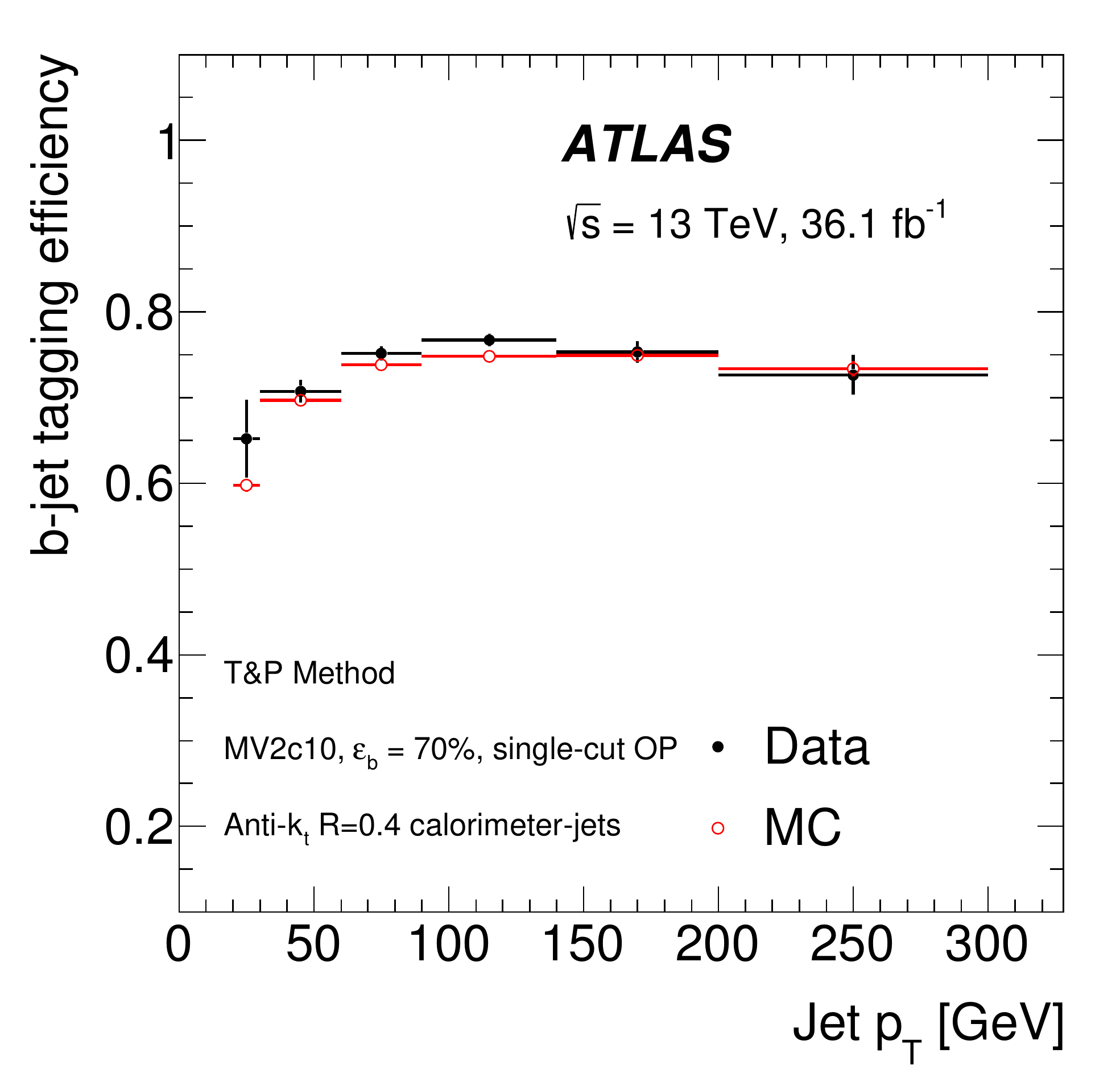}
}
\\
\subfloat[][]{
\includegraphics[width=0.49\textwidth]{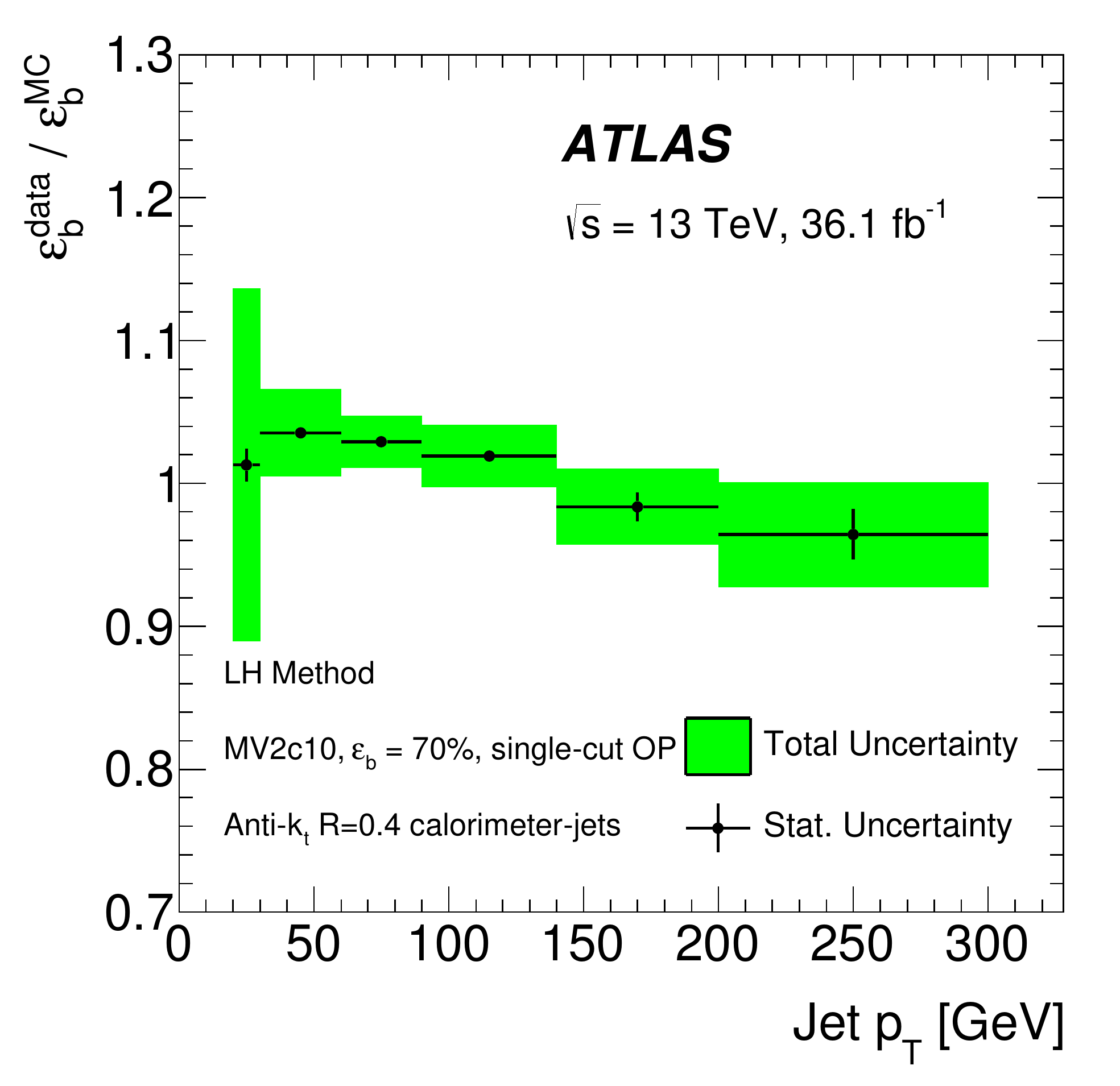}
}
\subfloat[][]{
\includegraphics[width=0.49\textwidth]{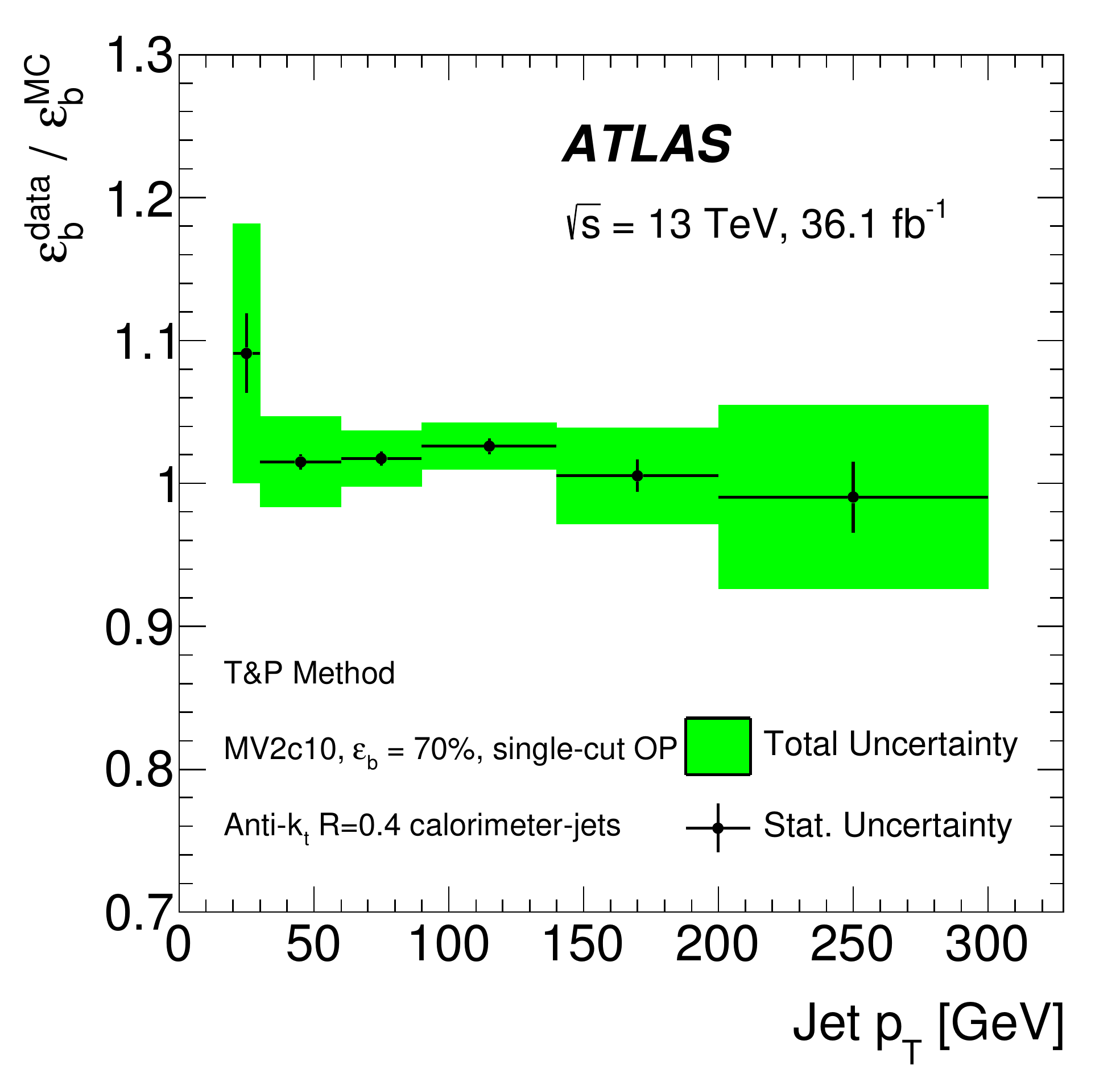}
}
\caption{Top: The \beff\ measured in data (full circles) and
  simulation (open circles), corresponding to the
  70\% \beff\ single-cut OP, as
a function of the jet \pt\ using (a) the LH method and (b) the T\&P method, for $R=0.4$ calorimeter-jets. The error bars correspond to the total statistical and systematic
uncertainties. Bottom: Data-to-simulation scale factors as a function of the
jet \pt\ using (c) the LH method and (d) the T\&P method. Both the statistical uncertainties (error bars) and total uncertainties (shaded region) are shown.}
\label{fig:ResEffSF}
\end{center}
\end{figure}
\begin{figure}
\begin{center}
\subfloat[][]{
\includegraphics[width=0.49\textwidth]{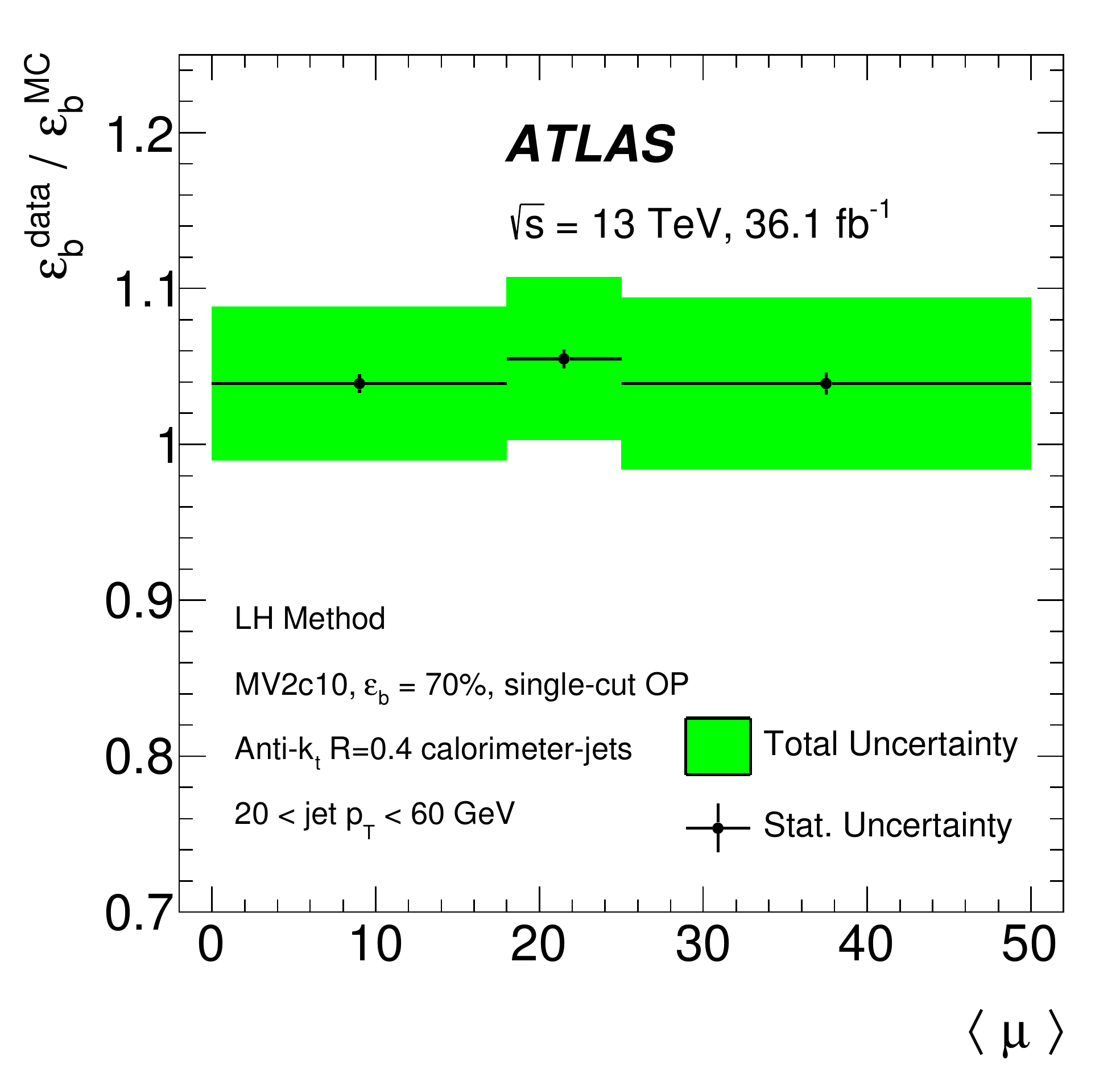}
}
\subfloat[][]{
\includegraphics[width=0.49\textwidth]{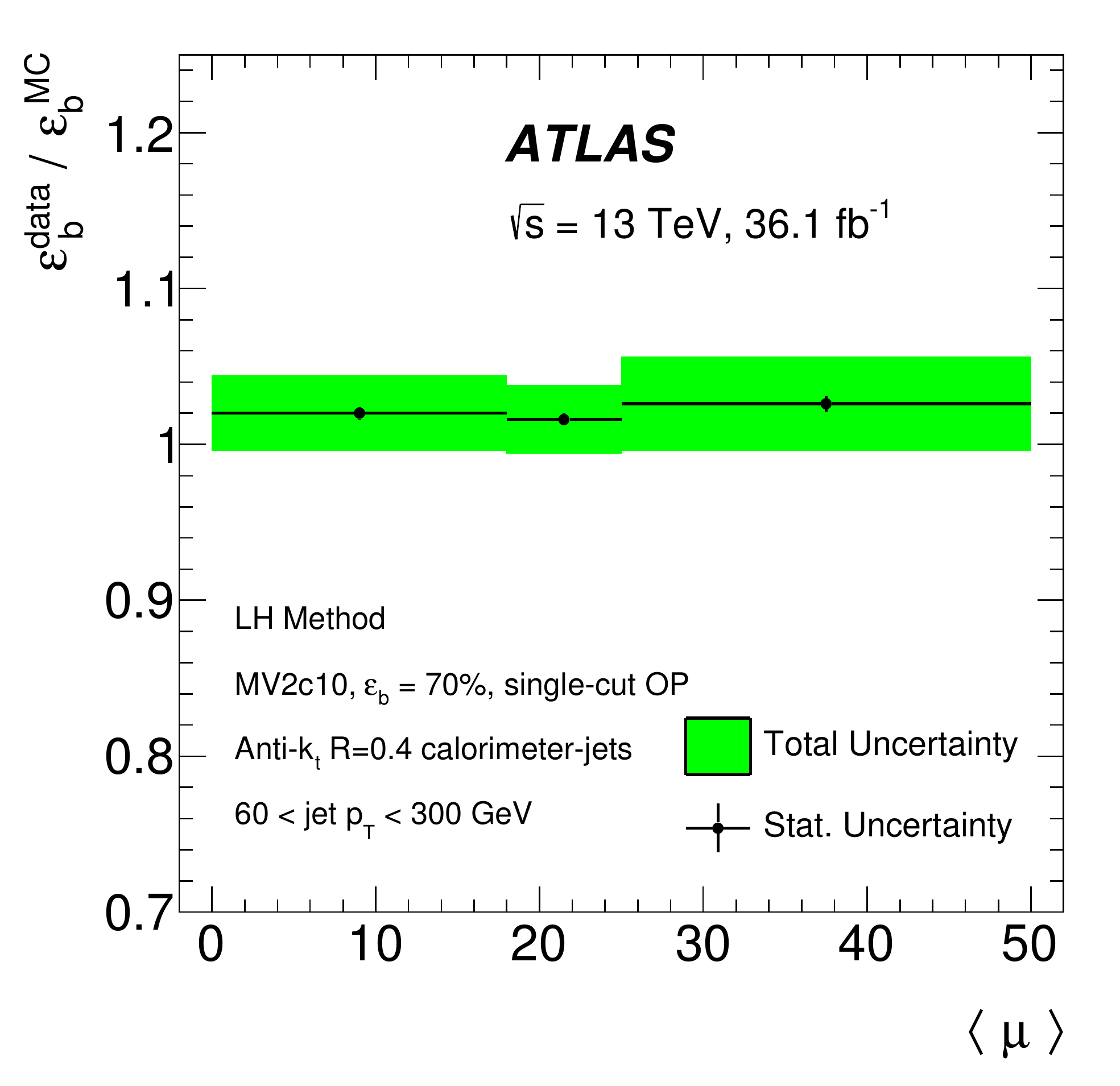}
}
\\
\subfloat[][]{
\includegraphics[width=0.49\textwidth]{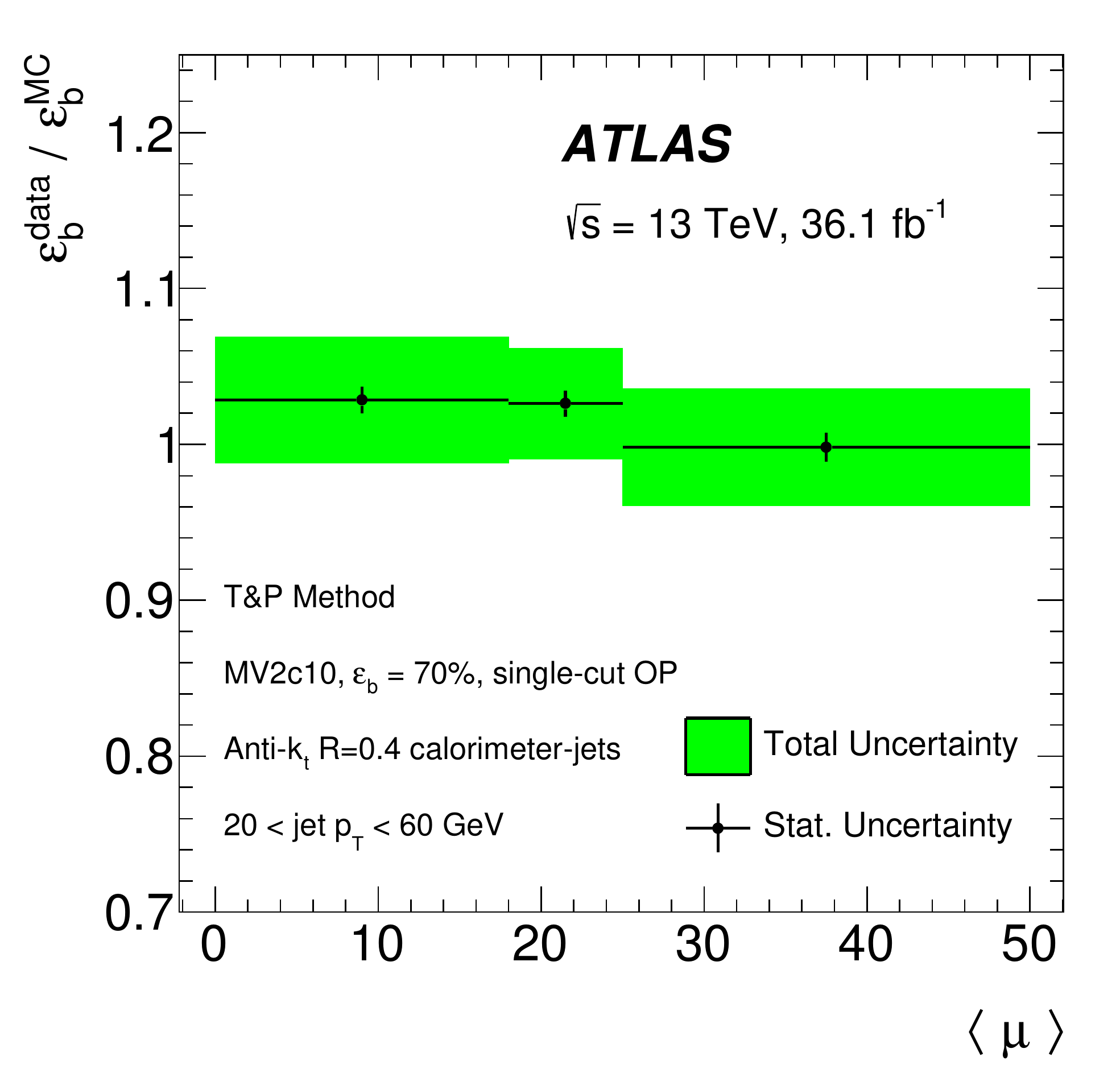}
}
\subfloat[][]{
\includegraphics[width=0.49\textwidth]{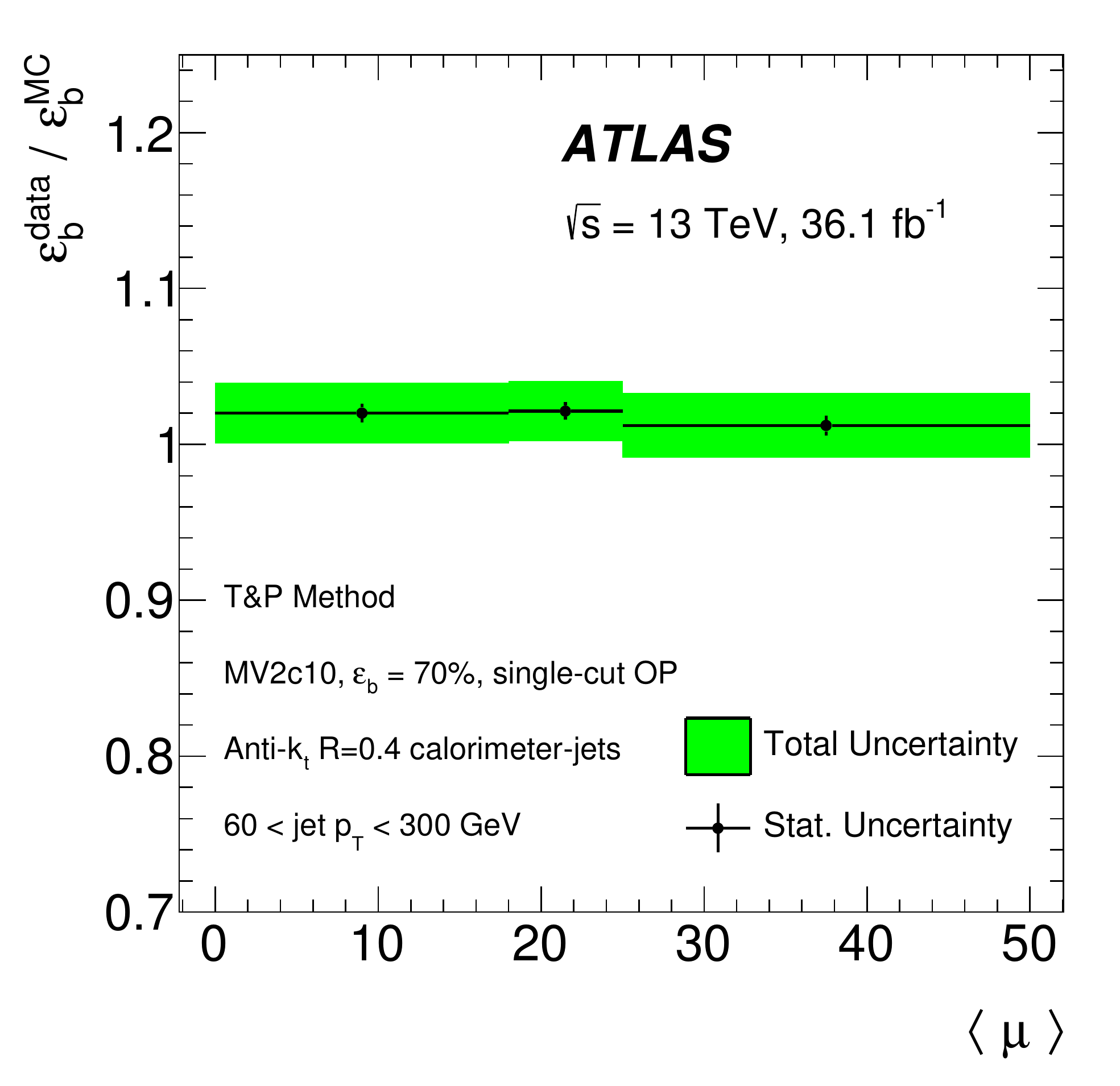}
}
\caption{Data-to-simulation scale factors, corresponding to the
  70\% \beff\ single-cut OP using $R=0.4$ calorimeter-jets, as a function of the
  average number of interactions per bunch crossing, $<\mu>$, for the LH method in the (a) 20 < \pT\ < 60~GeV region,  (b) 60 < \pT\ < 300~GeV region, and for the T\&P method in the (c) 20 < \pT\ < 60~GeV region, (d) 60 < \pT\ < 300~GeV region.  Both the statistical uncertainties (error bars) and total uncertainties (shaded region) are shown.}
\label{fig:ResEffSFPileup}
\end{center}
\end{figure}
\begin{figure}
\begin{center}
\subfloat[][]{
\includegraphics[width=0.49\textwidth]{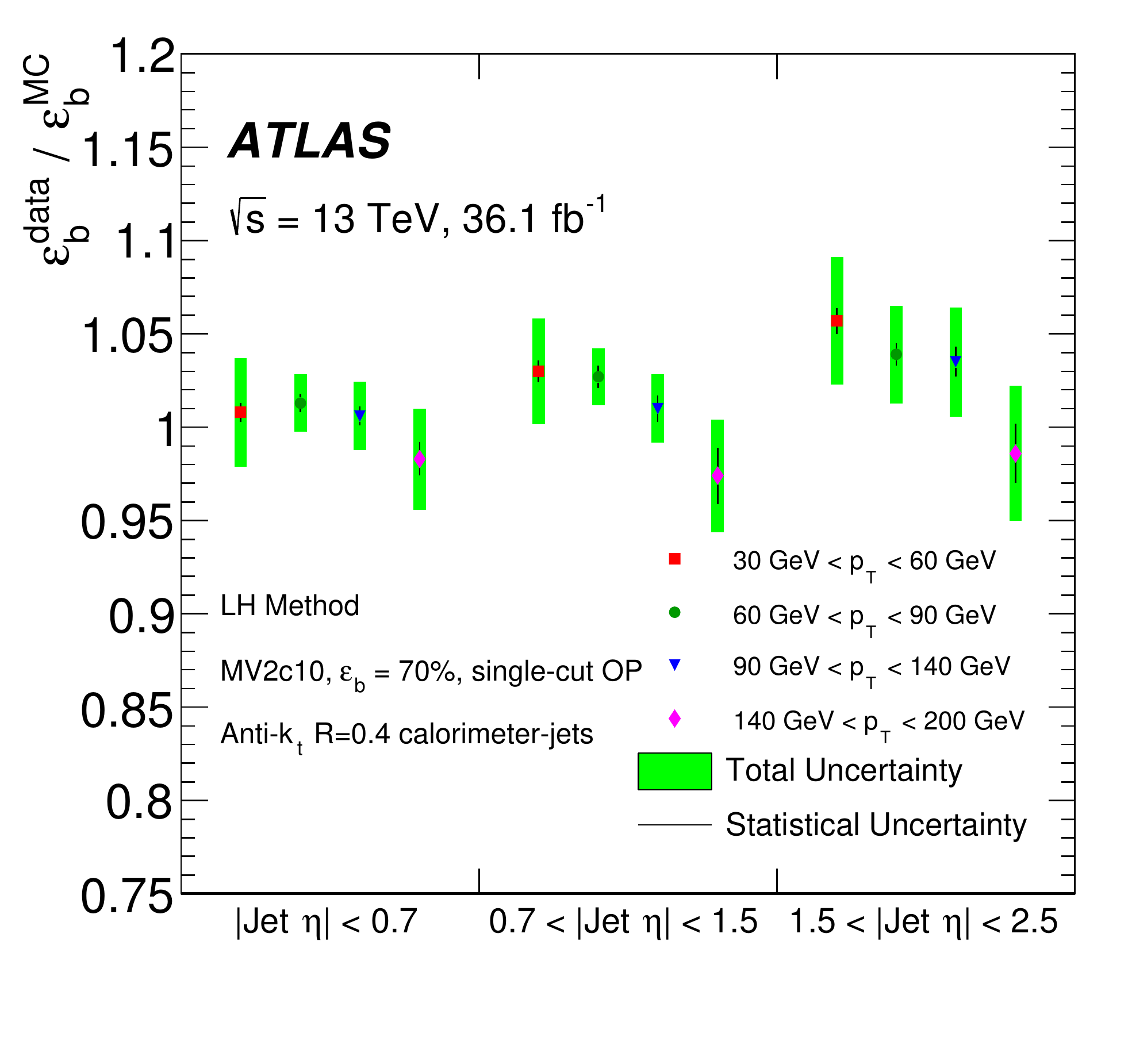}
}
\subfloat[][]{
\includegraphics[width=0.49\textwidth]{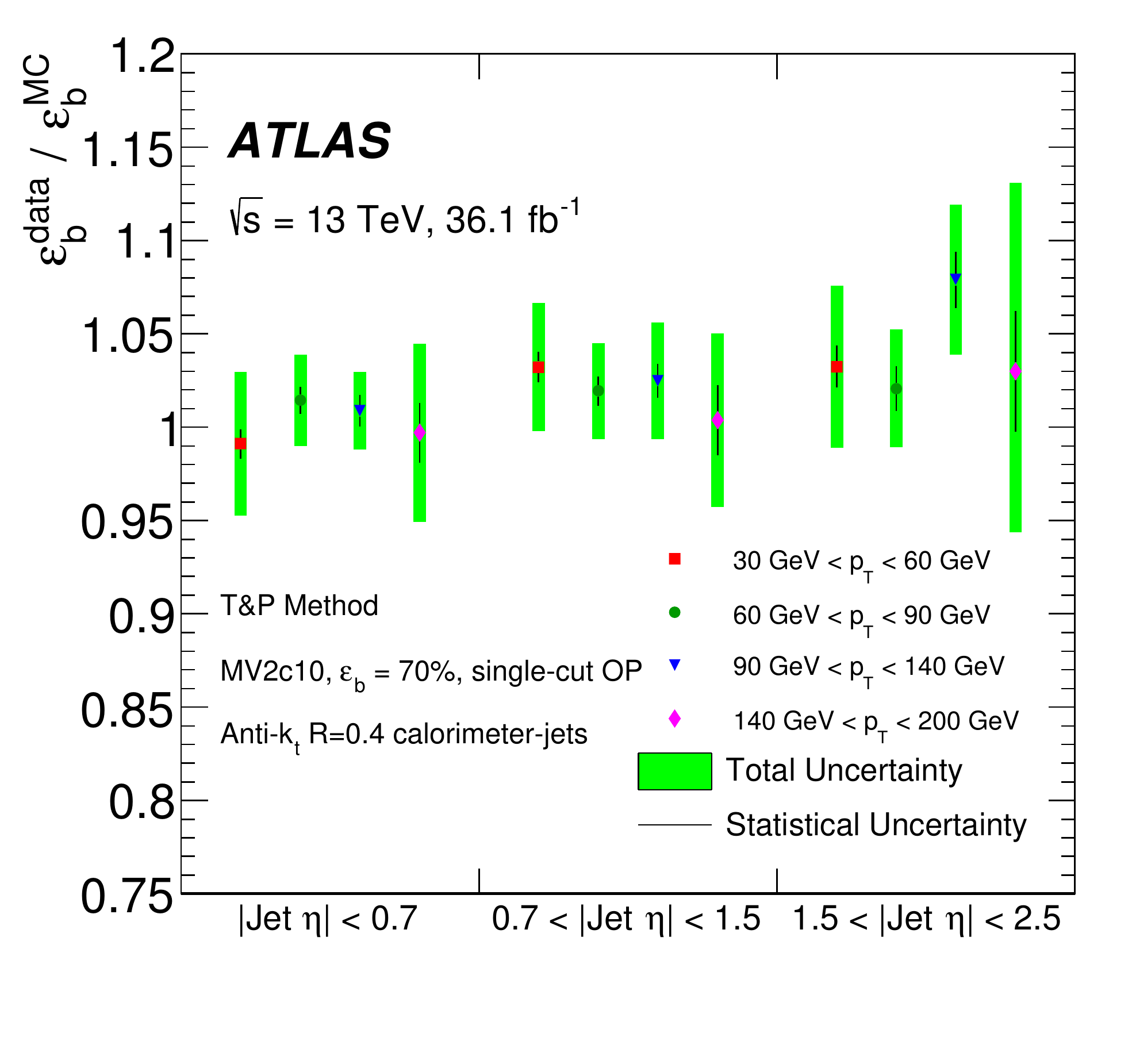}
}
\caption{Data-to-simulation scale factors, corresponding to the
  70\% \beff\ single-cut OP, as a function of the
  jet $|\eta|$, in (a) the LH method, and (b) the T\&P method, for $R=0.4$ calorimeter-jets. Both the statistical uncertainties (error bars) and total uncertainties (shaded region) are shown.}
\label{fig:ResEffSFEta}
\end{center}
\end{figure}

Figure~\ref{fig:ResEffSF} shows the measured efficiency in data
and simulation and the data-to-simulation scale factors as
a function of the jet \pt\ for both the T\&P and LH
methods, corresponding to the 70\% \beff\ single-cut OP, for $R=0.4$ calorimeter-jets. The efficiencies determined in simulation and
data agree within their uncertainties, resulting in scale
factors close to unity.
It can be seen that the resulting
data-to-simulation scale factors are in agreement between the two methods, with similar
central values and uncertainties. Scale factors were also measured as a
function of the average number of interactions per
bunch crossing, in selected \pt\ bins, and the jet $\eta$, using both the LH and T\&P methods, and are shown in Figures~\ref{fig:ResEffSFPileup} and~\ref{fig:ResEffSFEta}, respectively, for the single-cut OP.
The data-to-simulation scale factors are observed not to have a strong dependence on either variable.
The \bjet\ tagging efficiency in simulation varies by less than 1\% over the range $0 < \langle \mu \rangle < 50$, and by up to 5\% of the range $0 < |\eta| < 2.5$.

Tables~\ref{tab:PDFsys} and \ref{tab:TPsys} show the data-to-simulation scale factors, and the statistical, systematic and
total uncertainties separately for each \pt\ bin.
Depending on the \pt\ bin, the total uncertainties
range between 2\% and 12\% for the LH method and 2\% and 9\% for
the T\&P method, with the statistical uncertainty component ranging
between 0.3\% and 1.8\% for the LH method and 0.5\% and 2.8\% for the T\&P
method. A reduction in the statistical uncertainty is achieved in the LH method by combining measurements from four channels, as well as
exploiting the correlations in the events, while in the case of the
T\&P method only the $e\mu$ + 2-jet channel is used.
 The systematic uncertainty component varies from
1.5\% to 8.6\% depending on the jet \pt\ for the T\&P method, while in
the case of the LH method, the effect of systematic uncertainties ranges between 1.8\%
and 12\%.  However, in the LH method, the total uncertainty is smaller for a larger jet-\pT\ range, and this method is therefore used as the default $b$-jet calibration. The dominant sources of uncertainty in both methods relate to the modelling of the \ttbar\ sample and alter the predicted flavour composition, to which both methods are particularly sensitive. The
application of the sample-selection BDT reduces the impact of these
uncertainties by up to 50\% due to the increase of the \bjet\ purity and the removal of regions of phase space which have large modelling uncertainties. At very low and high jet \pt, the uncertainties related to
the measurement of the jet energy scale and resolution also become
significant. Normalisation and modelling of the \zjets\ background, as
well as the normalisation of the diboson backgrounds have a larger effect in
the LH method than in the T\&P method.  This is due to the inclusion of the events with three jets
and events with the same-flavour leptons in the final state, as these regions have a larger contribution from the \zjets\ and diboson backgrounds.

\begin{table}
\renewcommand{\arraystretch}{1.1} 
\centering
\scalebox{0.95}{
\begin{tabular}{l|c|c|c|c|c|c}
\hline\hline
\multicolumn{7}{c}{LH Method} \\
\hline\hline
\pT\ interval [GeV] & 20--30 & 30--60 & 60--90 & 90--140 & 140--200 & 200--300 \\
\hline\hline
Scale factor & 1.013 & 1.035 & 1.029 & 1.019 & 0.984 & 0.964 \\
Total uncertainty & 0.123 & 0.030 & 0.018 & 0.022 & 0.026 & 0.037 \\
Statistical uncertainty & 0.012 & 0.003 & 0.004 & 0.004 & 0.010 & 0.018 \\
Systematic uncertainty & 0.123 & 0.030 & 0.018 & 0.021 & 0.024 & 0.032 \\
\hline\hline
\multicolumn{7}{c}{Systematic Uncertainties [\%]} \\
\hline\hline
Matrix element modelling ($t\bar{t}$) & 3.2 & 0.3 & 0.9 & 1.1 & 1.1 & 0.7 \\
Parton shower / Hadronisation ($t\bar{t}$) & 9.0 & 1.5 & 0.3 & 1.0 & 1.4 & 2.2 \\
NNLO top \pT, $t\bar{t}$ \pT\ reweighting (\ttbar) & 0.1 & 0.1 & 0.1 & 0.3 & 0.5 & 0.9 \\
PDF reweighting ($t\bar{t}$) & 0.9 & 0.2 & 0.2 & 0.3 & 0.4 & 0.4 \\
More / less parton radiation ($t\bar{t}$) & 1.7 & 0.9 & 0.4 & 0.3 & 0.6 & 0.4 \\
Matrix element modelling (single top) & 0.5 & 0.2 & 0.2 & 0.2 & 0.3 & 0.1 \\
Parton shower / Hadronisation (single top) & 1.1 & 0.1 & 0.1 & 0.0 & 0.1 & 0.2 \\
More / less parton radiation (single top) & 0.0 & 0.0 & 0.0 & 0.1 & 0.1 & 0.1 \\
DR vs. DS (single top) & 0.1 & 0.1 & 0.1 & 0.1 & 0.1 & 0.2 \\
Modelling ($Z+$jets) & 0.6 & 0.5 & 0.5 & 0.9 & 0.6 & 1.2 \\
\pT\ reweighting ($Z+$jets) & 0.0 & 0.1 & 0.0 & 0.1 & 0.1 & 0.1 \\
\hline
MC non-closure & 1.2 & 0.0 & 0.0 & 0.0 & 0.0 & 0.0 \\
\hline
Normalisation single top & 0.2 & 0.1 & 0.0 & 0.1 & 0.1 & 0.1 \\
Normalisation $Z+$jets & 1.8 & 0.5 & 0.5 & 0.4 & 0.5 & 0.5 \\
Normalisation $Z+b/c$ & 0.4 & 0.1 & 0.1 & 0.0 & 0.0 & 0.0 \\
Normalisation diboson & 1.6 & 1.1 & 0.7 & 0.6 & 0.7 & 0.8 \\
Normalisation misid. leptons & 0.7 & 0.7 & 0.6 & 0.6 & 0.5 & 0.5 \\
\hline
Pile-up reweighting & 0.3 & 0.0 & 0.0 & 0.2 & 0.3 & 0.6 \\
Electron efficiency/resolution/scale/trigger & 0.1 & 0.0 & 0.0 & 0.0 & 0.0 & 0.0 \\
Muon efficiency/resolution/scale/trigger & 0.1 & 0.0 & 0.0 & 0.0 & 0.2 & 0.2 \\
\met & 0.1 & 0.0 & 0.0 & 0.0 & 0.1 & 0.1 \\
JVT & 0.1 & 0.0 & 0.1 & 0.1 & 0.0 & 0.1 \\
Jet energy scale (JES) & 6.8 & 1.4 & 0.5 & 0.4 & 0.5 & 0.7 \\
Jet energy resolution (JER) & 0.0 & 1.3 & 0.3 & 0.1 & 0.3 & 0.4 \\
Light-flavour jet mis-tag rate & 0.6 & 0.1 & 0.0 & 0.0 & 0.0 & 0.0 \\
$c$-jet mis-tag rate & 0.6 & 0.1 & 0.1 & 0.0 & 0.0 & 0.0 \\
Luminosity & 0.2 & 0.1 & 0.1 & 0.1 & 0.1 & 0.1 \\
\hline\hline
\end{tabular}
}
\caption{Data-to-simulation scale factors and associated
  uncertainties for the 70\% \beff\ single-cut operating point of the MV2c10
  $b$-jet tagging algorithm using the LH method, for $R=0.4$ calorimeter-jets, as a function of the jet \pt.}
\label{tab:PDFsys}
\end{table}

\begin{table}
\renewcommand{\arraystretch}{1.1} 
\centering
\scalebox{0.95}{
\begin{tabular}{l|c|c|c|c|c|c} 
\hline\hline
\multicolumn{7}{c}{T\&P Method} \\
\hline\hline

     \pt\ interval [GeV] & 20--30 & 30--60 & 60--90 & 90--140 & 140--200 & 200--300\\ 
\hline\hline
Scale factor & 1.091  & 1.015  & 1.017  & 1.026  & 1.005  & 0.990 \\
Total uncertainty & 0.091  & 0.032  & 0.019  & 0.016  & 0.034  & 0.064 \\
Statistical uncertainty & 0.028  & 0.005  & 0.005  & 0.006  & 0.011  & 0.025 \\
Systematic uncertainty & 0.086  & 0.031  & 0.019  & 0.015  & 0.032  & 0.059 \\
\hline\hline
\multicolumn{7}{c}{Systematic Uncertainties [\%]} \\
\hline\hline
Matrix element modelling (\ttbar) & 2.6  & 0.5  & 1.1  & 0.8  & 1.3  & 3.5 \\
Parton shower / Hadronisation (\ttbar) & 2.0  & 2.1  & 0.7  & 0.4  & 1.9  & 1.8 \\
NNLO top \pt, \ttbar\ \pt\ reweighting (\ttbar) & 0.2  & 0.2  & 0.1  & 0.1  & 0.1  & 0.2 \\
PDF reweighting (\ttbar) & 2.9  & 0.6  & 0.6  & 0.7  & 1.3  & 2.2 \\
More / less parton radiation (\ttbar) & 3.8  & 1.7  & 0.8  & 0.7  & 0.7  & 2.8 \\
Matrix element modelling (single top) & 1.1  & 0.2  & 0.3  & 0.2  & 0.5  & 0.9 \\
Parton shower / Hadronisation (single top) & 1.1  & 0.2  & 0.2  & 0.2  & 0.5  & 0.9 \\
More / less parton radiation (single top) & 1.1  & 0.2  & 0.2  & 0.3  & 0.5  & 1.0 \\
DR vs. DS (single top) & 1.1  & 0.2  & 0.2  & 0.2  & 0.5  & 1.0 \\
Modelling ($Z+$jets) & 1.1  & 0.2  & 0.2  & 0.2  & 0.5  & 0.9 \\
\hline
Limited size of simulated samples & 1.6  & 0.3  & 0.3  & 0.3  & 0.6  & 1.4 \\
\hline
Normalisation single top & 0.3  & 0.1  & 0.1  & 0.0  & 0.0  & 0.0 \\
Normalisation $Z+$jets & 0.0  & 0.0  & 0.0  & 0.0  & 0.0  & 0.0 \\
Normalisation $Z+b/c$ & 0.1  & 0.1  & 0.1  & 0.0  & 0.1  & 0.0 \\
Normalisation diboson & 0.1  & 0.1  & 0.0  & 0.0  & 0.0  & 0.0 \\
Normalisation misid. leptons & 0.3  & 0.1  & 0.0  & 0.1  & 0.1  & 0.1 \\
\hline
Pile-up reweighting & 1.9  & 0.4  & 0.4  & 0.2  & 0.7  & 0.9 \\
Electron efficiency/resolution/scale/trigger & 0.0  & 0.0  & 0.0  & 0.0  & 0.1  & 0.0 \\
Muon efficiency/resolution/scale/trigger  & 0.3  & 0.0  & 0.0  & 0.0  & 0.0  & 0.0 \\
\met & 0.2  & 0.0  & 0.0  & 0.0  & 0.0  & 0.0 \\
JVT & 0.1  & 0.0  & 0.0  & 0.0  & 0.0  & 0.0 \\
Jet energy scale (JES) & 3.8  & 0.7  & 0.3  & 0.3  & 0.7  & 0.3 \\
Jet energy resolution (JER) & 1.5  & 0.5  & 0.2  & 0.1  & 0.2  & 0.7 \\
Light-flavour jet mis-tag rate & 0.2  & 0.0  & 0.1  & 0.0  & 0.1  & 0.0 \\
$c$-jet mis-tag rate & 0.1  & 0.0  & 0.0  & 0.0  & 0.0  & 0.0 \\
Luminosity & 0.2  & 0.1  & 0.0  & 0.0  & 0.0  & 0.0 \\
\hline\hline
\end{tabular}

}
\caption{Data-to-simulation scale factors and associated
  uncertainties for the 70\% \beff\ single-cut operating point of the MV2c10
  $b$-jet tagging algorithm using the T\&P method, for $R=0.4$ calorimeter-jets, as a function of the jet \pt.}
\label{tab:TPsys}
\end{table}

An additional uncertainty is included to extrapolate the measured uncertainties to higher jet \pT, which is not measured here but is of interest in some physics analyses.
This term is calculated from simulated events by considering variations of the quantities affecting the $b$-tagging performance such as the impact parameter resolution, percentage of poorly measured tracks, description of the detector material, and the track multiplicity per jet. The dominant effect on the uncertainty when extrapolating at high \pT\ is related to the different tagging efficiencies after smearing the tracks' impact parameters according to the resolutions measured in data and simulation. The difference in the impact parameter resolution is due to effects from alignment, dead modules and additional material not properly modelled in the simulation. The impact of the $b$-tagging efficiency uncertainty increases with jet \pT\ and reaches 15\% above $1.5~\TeV$. 

\begin{figure}[h]
\begin{center}
\subfloat[][]{
\includegraphics[width=0.49\textwidth]{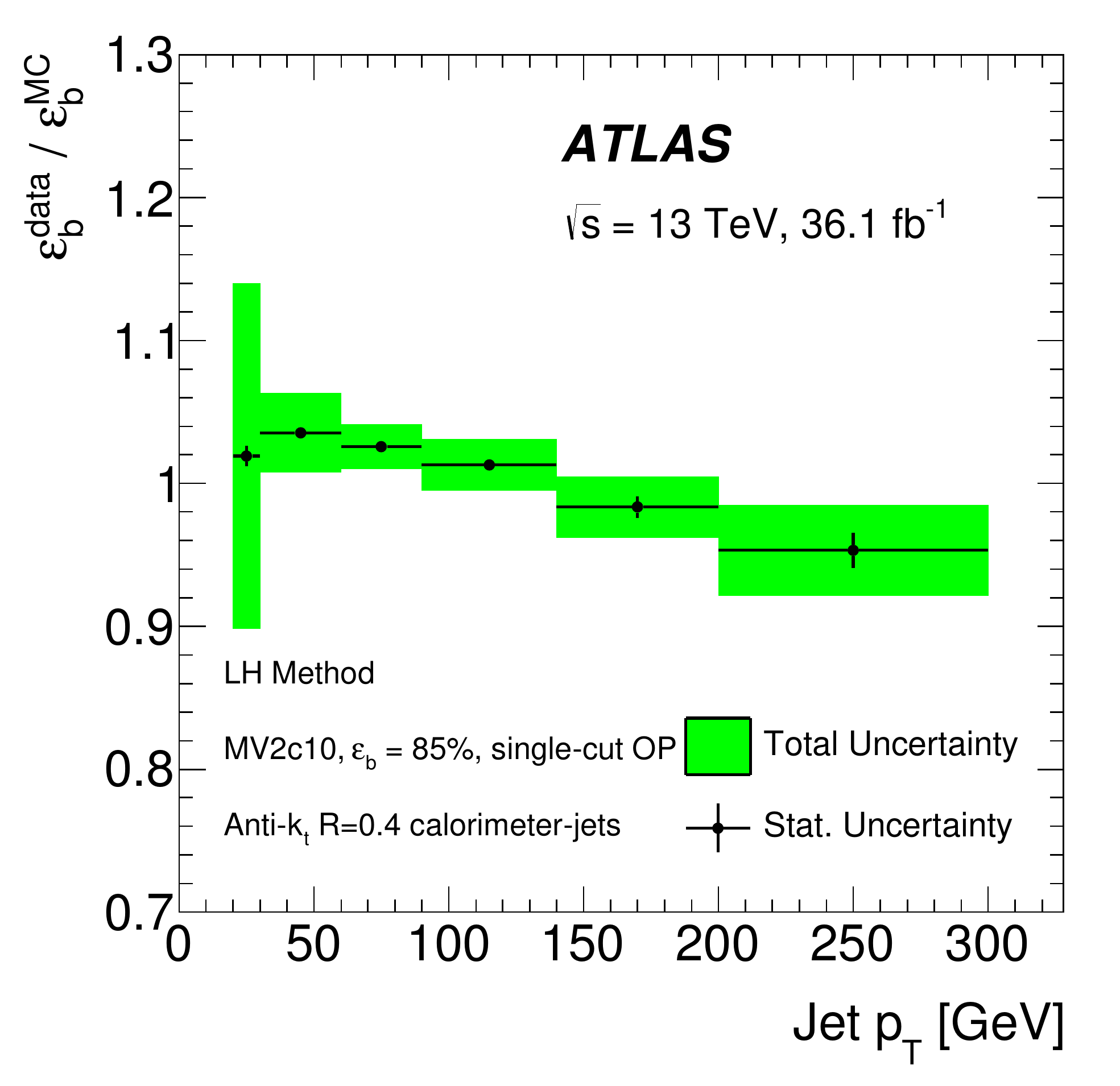}
}
\subfloat[][]{
\includegraphics[width=0.49\textwidth]{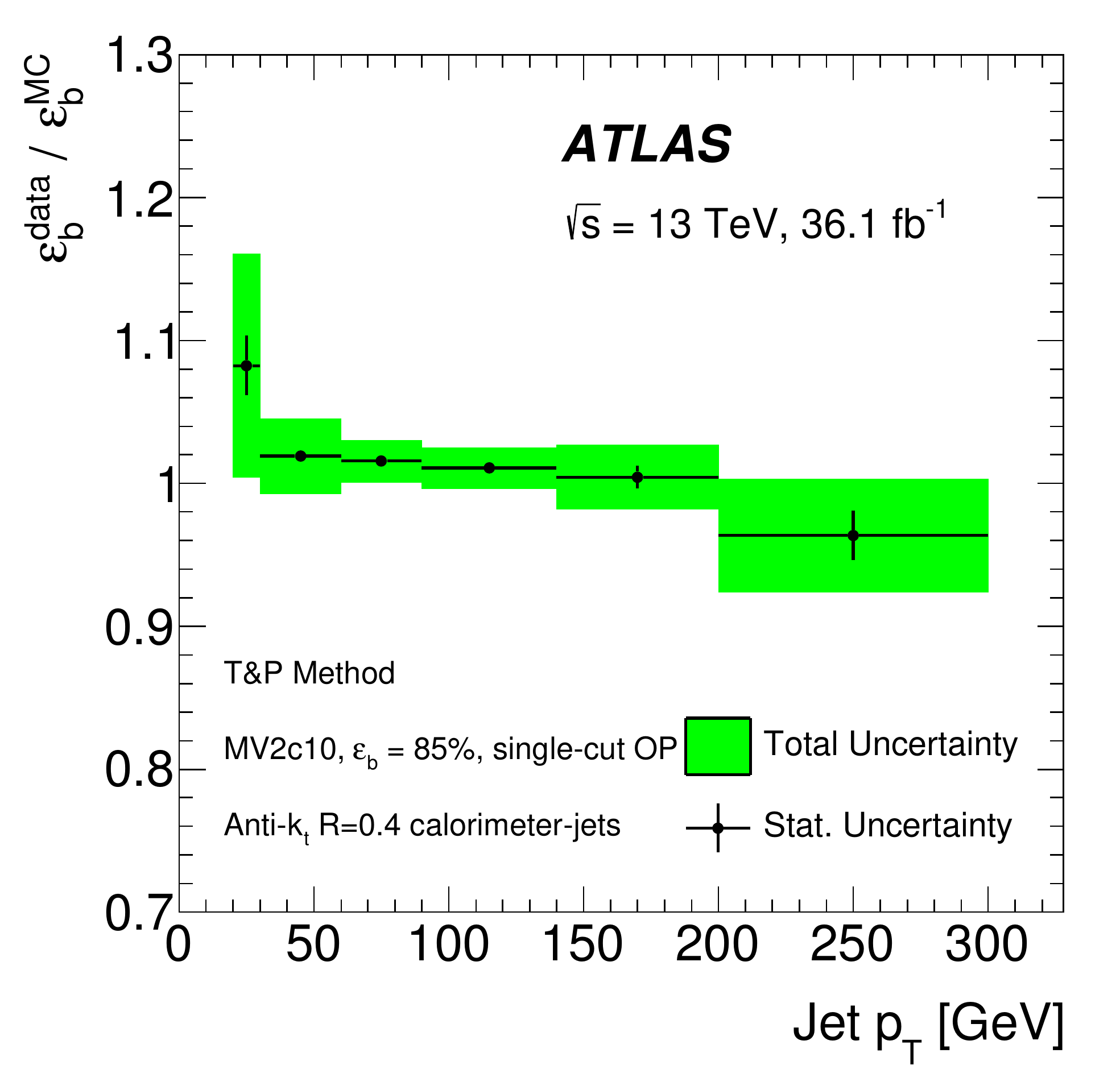}
}
\caption{
Data-to-simulation scale factors, corresponding to the
  85\% \beff\ single-cut OP, as a function of the
  jet \pT, in (a) the LH method, and (b) the T\&P method, for $R=0.4$ calorimeter-jets. Both the statistical uncertainties (error bars) and total uncertainties (shaded region) are shown.}
\label{fig:PDFEffSF85EFF}
\end{center}
\end{figure}

Similar measurements were conducted for other types of
reconstructed jets and other $b$-tagging OPs used within the ATLAS physics program, and are used accordingly. 
As an additional example, Figure~\ref{fig:PDFEffSF85EFF} presents the measured data-to-simulation scale factors as
a function of the jet \pt\ for both the T\&P and LH
methods, corresponding to the 85\% \beff\ single-cut OP for $R=0.4$ calorimeter-jets.

\begin{figure}[h]
\begin{center}
\subfloat[][]{
\includegraphics[width=0.49\textwidth]{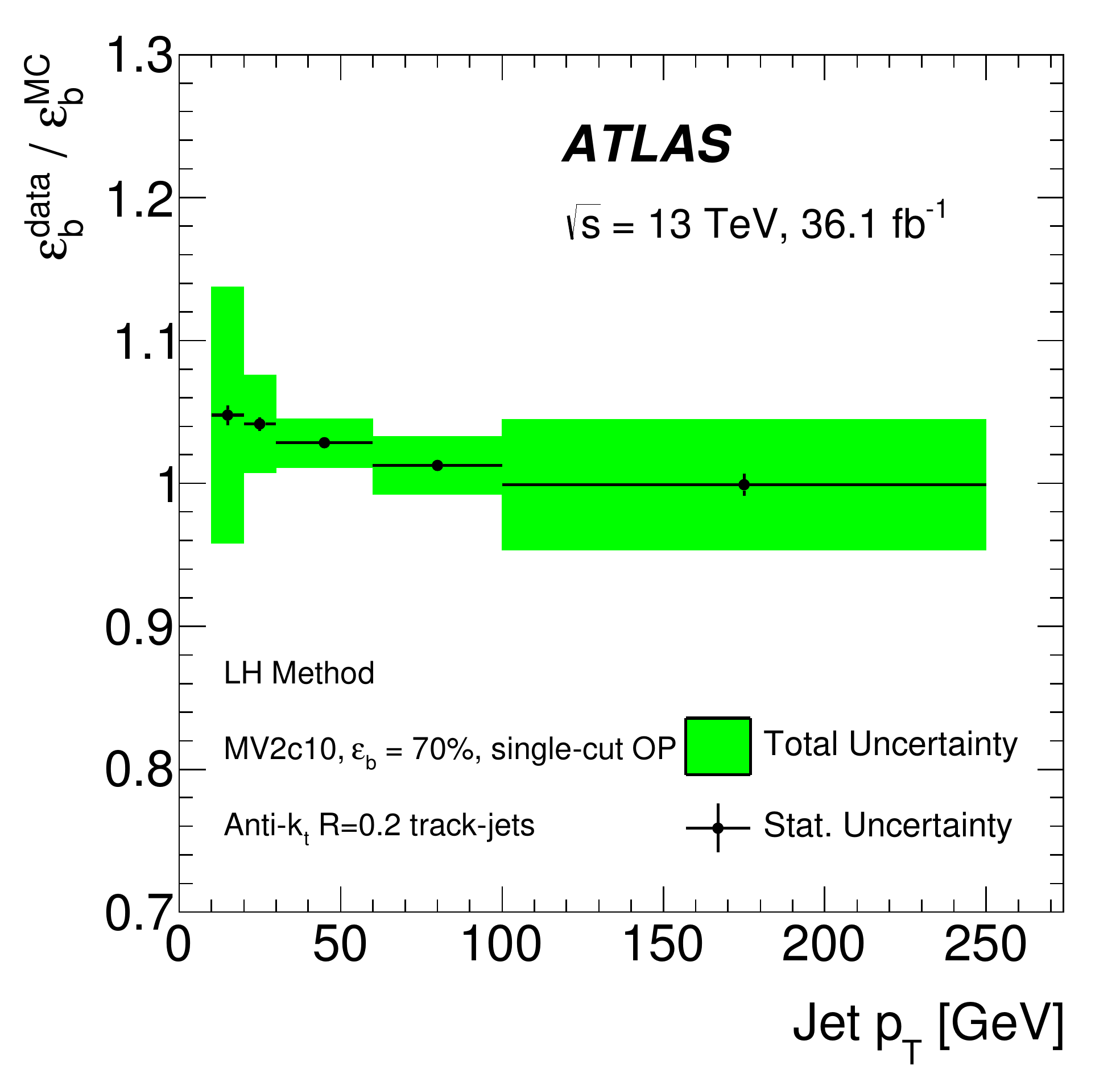}
}
\subfloat[][]{
\includegraphics[width=0.49\textwidth]{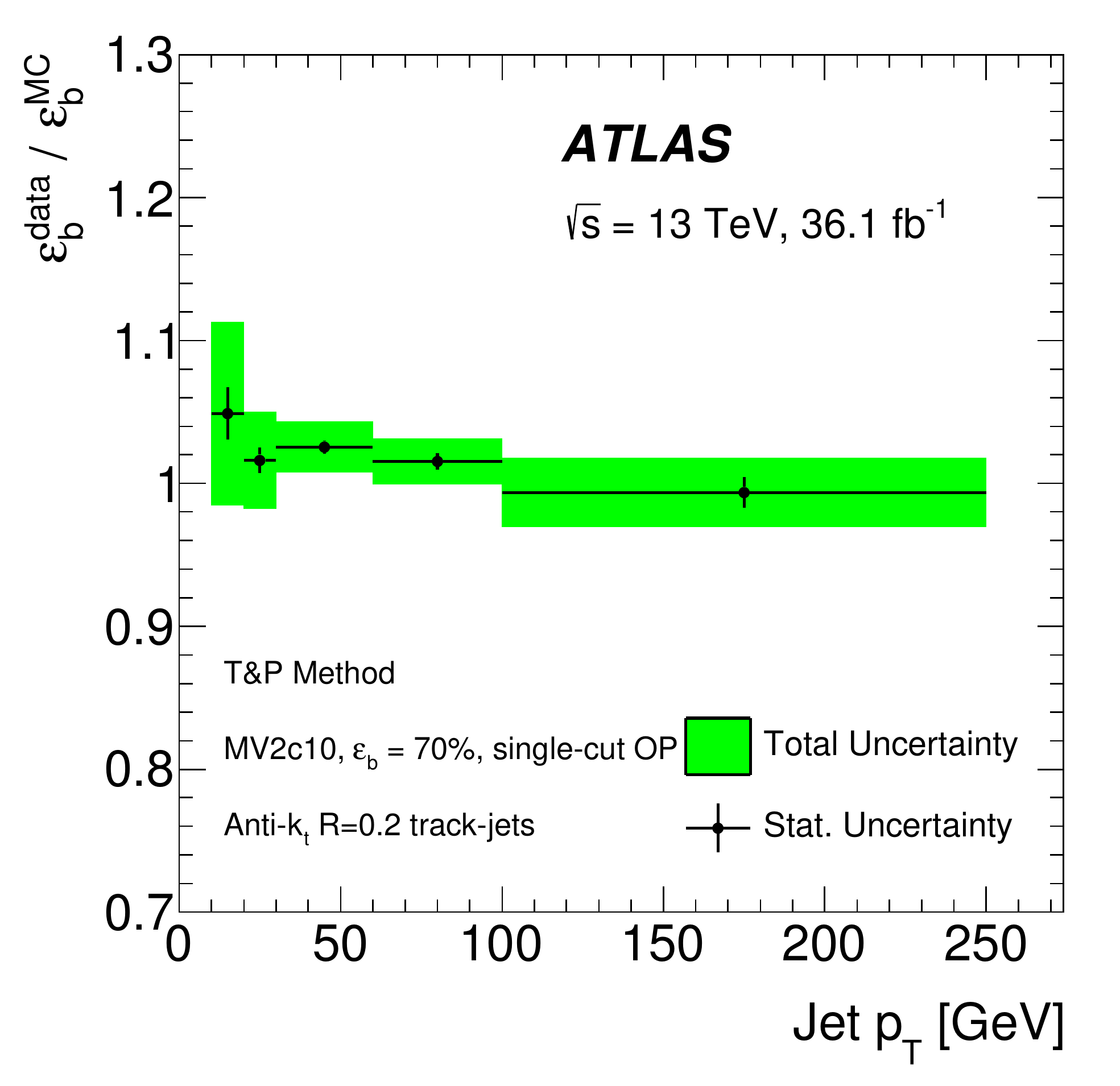}
}
\caption{
Data-to-simulation scale factors as a function of
the jet \pt\ using (a) the LH method and (b) the T\&P method for $R=0.2$ track-jets. Both the statistical uncertainties (error bars) and total uncertainties (shaded region) are shown. The results correspond to the 70\% \beff\ single-cut operating point of the MV2c10 $b$-tagging algorithm.}
\label{fig:PDFEffSFtrack}
\end{center}
\end{figure}

Scale factors for the anti-$k_{t}$ $R=0.2$ track-jets at the 70\% single-cut OP are presented in Figure~\ref{fig:PDFEffSFtrack}.
The efficiencies determined in simulation and
data agree within their uncertainties, resulting in scale
factors close to unity.

\label{labEVandSmooth}
\subsection{Generator dependence of the efficiency scale factors}

The use of \textsc{EvtGen} ensures that \textsc{Pythia} and \textsc{Herwig} use a consistent lifetime and decay model for all $b$-hadron species (e.g. $B^{+}$, $B^{0}$, $B^{0}_{s}$), which reduces the differences between the $b$-jet tagging efficiencies predicted by the two generators. Nevertheless, the intrinsic tagging efficiency of a $b$-jet still depends on several aspects which are not harmonised between the different generators, such as: the initial production fractions of the different $b$-hadron species, the fragmentation function, the number of additional charged particles not from the $b$-hadron in the jet and the relative topology of the $b$-hadron and the jet. These differences cause the intrinsic $b$-jet tagging efficiency of a sample
 to vary depending on the hadronisation/fragmentation generator. Therefore, when using a simulated sample with a different fragmentation model to that used to derive the data-to-simulation scale factors (i.e. \textsc{Powheg+Pythia6}), it is necessary to include additional  generator-dependent scale factors.
Generator-dependent data-to-simulation scale factors are determined as the ratio of the predicted $b$-jet tagging efficiencies 
in each jet for the generator in question and the reference of \textsc{Powheg}+\textsc{Pythia6},
with the scale differing from 1 by less than 5\% for $b$-jets.

\subsection{Smoothing of the efficiency scale factors}
\label{smooth}

For use in physics measurements, the data-to-simulation efficiency scale factors are smoothed from the initial six bins in jet~\pt\ using a local polynomial kernel estimator~\cite{wand1995kernel}. This procedure is performed in order to avoid any boundary effects, and to prevent distortions in the distributions of interest in analyses when applying the scale factors.

\begin{figure}[h]
\begin{center}
\includegraphics[width=.60\textwidth]{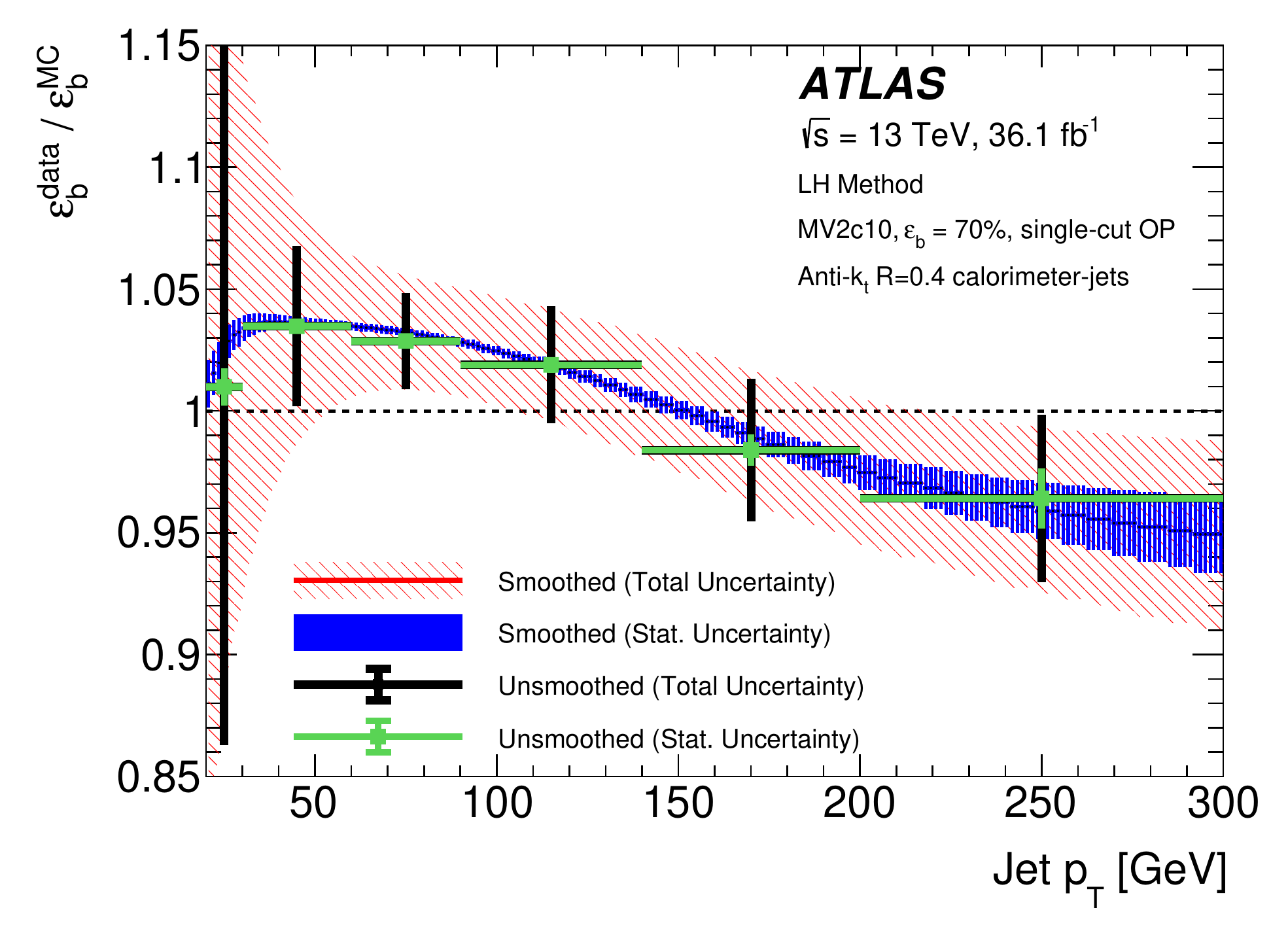}
\caption{A comparison of the data-to-simulation scale factors before and after smoothing is applied for the 70\% \beff\ single-cut operating point of the MV2c10 $b$-tagging algorithm, for $R=0.4$ calorimeter-jets.  The scale factors have been measured using the LH method.
The total and statistical uncertainties before applying smoothing are represented by error bars, while the total  and statistical uncertainties after applying smoothing correspond to the filled area.
}
\label{fig:smoothedSFs}
\end{center}
\end{figure}

The result of the smoothing of the \bjet\ scale factor for the 70\% operating point in the LH method is shown in Figure~\ref{fig:smoothedSFs}. All the per-bin systematic uncertainties are added in quadrature and shown together with the statistical uncertainties for the calibrated bins of jet \pt.

\subsection{Reduction of the nuisance parameters}

The total uncertainties in the data-to-simulation efficiency scale factors presented in Section~\ref{sec:results} are calculated as a sum in quadrature of the statistical uncertainty and individual components  of the systematic uncertainty. However, for the application to physics analyses, a statistically more correct approach based on varying each source of uncertainty by $\pm1\sigma$, independently, and considering its effect on the data-to-simulation efficiency scale factors in each bin, gives a more accurate estimate of the effect of the $b$-tagging uncertainty on the result.
If done in this way, a large number of uncertainties (one per source) would need to be taken into account.
Thus, reducing the number of systematic uncertainties that need to be considered, while still conserving the correct dependence on the jet \pt\ and jet $\eta$, is beneficial.

A method for reducing the number of systematic uncertainties while preserving the bin-to-bin correlations was developed, and is based on an eigenvalue decomposition of the covariance matrix of systematic and statistical variations. It starts from the construction of the $6\times 6$ covariance matrix corresponding to each source of uncertainty in the six bins of jet~\pt\ used for the calibration. Since bin-to-bin correlations are assumed, these matrices have non-zero off-diagonal elements. The total covariance matrix is constructed by summing these covariance matrices corresponding to different sources of uncertainty. As the total covariance matrix is a symmetric, positive-definite matrix, an eigenvector decomposition can be performed. 
Such a procedure provides orthogonal variations whose size is given by the square root of the corresponding eigenvalues. The resulting number of
variations is six, corresponding to the number of bins used for the calibration, and is an important simplification in the implementation of systematic uncertainties in physics analyses.

Finally, most of the eigenvalue variations are very small and can be neglected without impacting the correlations or total uncertainty. The remaining eigenvalue variations can be further reduced by removing eigenvalue variations below a chosen threshold. However, preservation of the correlations comes at a cost, with some of the total uncertainty incorrectly removed. Thus, a tradeoff is made as to how much of the total uncertainty is preserved versus the correlations. Three different schemes of eigen-variation reduction are implemented: `loose' provides a complete description of the total uncertainty and correlations, `medium' has a small amount of loss in  the total uncertainty or correlation loss (of the order of 3\% relative difference), and `tight' has a more aggressive reduction, where more loss in the total uncertainty or correlation is tolerated  (of the order of 10--50\% relative difference).


\section{Conclusion}
\label{sec:conclusion}
The $b$-jet tagging efficiency of the ATLAS $b$-tagging algorithm has been measured using a high-purity sample of dileptonic \ttbar\ events selected from the 36.1~fb$^{-1}$ of data collected by the ATLAS detector in 2015 and 2016 from proton--proton collisions at a centre-of-mass energy
$\sqrt{s}=13$~TeV at the LHC. A boosted decision tree, based on event topology only, is used to select events in which two \bjets\ are present, reducing the contamination from events in which only one \bjet\ is reconstructed in the detector acceptance. The implementation of a boosted decision tree in the event selection reduces the dominant uncertainty in the modelling of the flavour of the jets in the \ttbar\ events by up to 50\%. Two methods are used to extract the efficiency from the \ttbar\ events, a tag-and-probe method and a combinatorial likelihood approach. The efficiency is extracted for $R=0.4$ calorimeter-jets in a transverse momentum range from 20 to 300~\gev, with data-to-simulation scale factors calculated by comparing the efficiency extracted from collision data to that obtained from simulation. The two methods produce consistent results with similar precision. The measured data-to-simulation scale factors are close to unity with a total uncertainty ranging from 2\% to 12\%. In addition, the data-to-simulation scale factors are measured as a function of the jet $\eta$ and the average number of interactions per bunch crossing, in selected bins of the jet \pt, and are found not to have a significant dependence on either of these variables.  The generator dependence of the data-to-simulation scale factors is assessed, along with procedures for smoothing the scale factors, and reducing the number of nuisance parameters arising from the data-to-simulation scale factors.


\section*{Acknowledgements}


We thank CERN for the very successful operation of the LHC, as well as the
support staff from our institutions without whom ATLAS could not be
operated efficiently.

We acknowledge the support of ANPCyT, Argentina; YerPhI, Armenia; ARC, Australia; BMWFW and FWF, Austria; ANAS, Azerbaijan; SSTC, Belarus; CNPq and FAPESP, Brazil; NSERC, NRC and CFI, Canada; CERN; CONICYT, Chile; CAS, MOST and NSFC, China; COLCIENCIAS, Colombia; MSMT CR, MPO CR and VSC CR, Czech Republic; DNRF and DNSRC, Denmark; IN2P3-CNRS, CEA-DRF/IRFU, France; SRNSFG, Georgia; BMBF, HGF, and MPG, Germany; GSRT, Greece; RGC, Hong Kong SAR, China; ISF, I-CORE and Benoziyo Center, Israel; INFN, Italy; MEXT and JSPS, Japan; CNRST, Morocco; NWO, Netherlands; RCN, Norway; MNiSW and NCN, Poland; FCT, Portugal; MNE/IFA, Romania; MES of Russia and NRC KI, Russian Federation; JINR; MESTD, Serbia; MSSR, Slovakia; ARRS and MIZ\v{S}, Slovenia; DST/NRF, South Africa; MINECO, Spain; SRC and Wallenberg Foundation, Sweden; SERI, SNSF and Cantons of Bern and Geneva, Switzerland; MOST, Taiwan; TAEK, Turkey; STFC, United Kingdom; DOE and NSF, United States of America. In addition, individual groups and members have received support from BCKDF, the Canada Council, CANARIE, CRC, Compute Canada, FQRNT, and the Ontario Innovation Trust, Canada; EPLANET, ERC, ERDF, FP7, Horizon 2020 and Marie Sk{\l}odowska-Curie Actions, European Union; Investissements d'Avenir Labex and Idex, ANR, R{\'e}gion Auvergne and Fondation Partager le Savoir, France; DFG and AvH Foundation, Germany; Herakleitos, Thales and Aristeia programmes co-financed by EU-ESF and the Greek NSRF; BSF, GIF and Minerva, Israel; BRF, Norway; CERCA Programme Generalitat de Catalunya, Generalitat Valenciana, Spain; the Royal Society and Leverhulme Trust, United Kingdom.

The crucial computing support from all WLCG partners is acknowledged gratefully, in particular from CERN, the ATLAS Tier-1 facilities at TRIUMF (Canada), NDGF (Denmark, Norway, Sweden), CC-IN2P3 (France), KIT/GridKA (Germany), INFN-CNAF (Italy), NL-T1 (Netherlands), PIC (Spain), ASGC (Taiwan), RAL (UK) and BNL (USA), the Tier-2 facilities worldwide and large non-WLCG resource providers. Major contributors of computing resources are listed in Ref.~\cite{ATL-GEN-PUB-2016-002}.

\clearpage
\printbibliography
\clearpage 
 
\begin{flushleft}
{\Large The ATLAS Collaboration}

\bigskip

M.~Aaboud$^\textrm{\scriptsize 34d}$,    
G.~Aad$^\textrm{\scriptsize 99}$,    
B.~Abbott$^\textrm{\scriptsize 124}$,    
O.~Abdinov$^\textrm{\scriptsize 13,*}$,    
B.~Abeloos$^\textrm{\scriptsize 128}$,    
D.K.~Abhayasinghe$^\textrm{\scriptsize 91}$,    
S.H.~Abidi$^\textrm{\scriptsize 164}$,    
O.S.~AbouZeid$^\textrm{\scriptsize 143}$,    
N.L.~Abraham$^\textrm{\scriptsize 153}$,    
H.~Abramowicz$^\textrm{\scriptsize 158}$
% ...
\bigskip
\\

% ...
$^{13}$Institute of Physics, Azerbaijan Academy of Sciences, Baku; Azerbaijan.\\
% ...
$^{34}$$^{(a)}$Facult\'e des Sciences Ain Chock, R\'eseau Universitaire de Physique des Hautes Energies - Universit\'e Hassan II, Casablanca;$^{(b)}$Centre National de l'Energie des Sciences Techniques Nucleaires (CNESTEN), Rabat;$^{(c)}$Facult\'e des Sciences Semlalia, Universit\'e Cadi Ayyad, LPHEA-Marrakech;$^{(d)}$Facult\'e des Sciences, Universit\'e Mohamed Premier and LPTPM, Oujda;$^{(e)}$Facult\'e des sciences, Universit\'e Mohammed V, Rabat; Morocco.\\
% ...
$^{91}$Department of Physics, Royal Holloway University of London, Egham; United Kingdom.\\
% ...
$^{99}$CPPM, Aix-Marseille Universit\'e, CNRS/IN2P3, Marseille; France.\\
% ...
$^{124}$Homer L. Dodge Department of Physics and Astronomy, University of Oklahoma, Norman OK; United States of America.\\
% ...
$^{128}$LAL, Universit\'e Paris-Sud, CNRS/IN2P3, Universit\'e Paris-Saclay, Orsay; France.\\
% ...
$^{143}$Santa Cruz Institute for Particle Physics, University of California Santa Cruz, Santa Cruz CA; United States of America.\\
% ...
$^{153}$Department of Physics and Astronomy, University of Sussex, Brighton; United Kingdom.\\
% ...
$^{158}$Raymond and Beverly Sackler School of Physics and Astronomy, Tel Aviv University, Tel Aviv; Israel.\\
% ...
$^{164}$Department of Physics, University of Toronto, Toronto ON; Canada.\\
% ...
$^{a}$ Also at  Department of Physics, University of Malaya, Kuala Lumpur; Malaysia.\\
$^{b}$ Also at Borough of Manhattan Community College, City University of New York, NY; United States of America.\\
$^{c}$ Also at Centre for High Performance Computing, CSIR Campus, Rosebank, Cape Town; South Africa.\\
$^{d}$ Also at CERN, Geneva; Switzerland.\\
$^{e}$ Also at CPPM, Aix-Marseille Universit\'e, CNRS/IN2P3, Marseille; France.\\
% ...
$^{*}$ Deceased

\end{flushleft}




\end{document}